\documentclass[pdflatex,sn-mathphys-num]{sn-jnl}% Math and Physical Sciences Numbered Reference Style

\usepackage{graphicx}%
\usepackage{multirow}%
\usepackage{amsmath,amssymb,amsfonts}%
\usepackage{amsthm}%
\usepackage{mathrsfs}%
\usepackage[title]{appendix}%
\usepackage{xcolor}%
\usepackage{textcomp}%
\usepackage{manyfoot}%
\usepackage{booktabs}%
\usepackage{algorithm}%
\usepackage{algorithmicx}%
\usepackage{algpseudocode}%
\usepackage{listings}%

\newcommand\degr{\mbox{$^\circ$}}% 

\begin{document}

\title{Simulation of non X-ray background for the DIffuse X-ray Explorer (DIXE) mission}

\author[1]{\fnm{Ruixuan} \sur{Tian}} \email{trx23@mails.tsinghua.edu.cn} 
\author[1]{\fnm{Junjie} \sur{Mao}}  \email{jmao@tsinghua.edu.cn}
\author[1]{\fnm{Jiejia} \sur{Liu}}  \email{liujj21@mails.tsinghua.edu.cn}
\author[1]{\fnm{Hai} \sur{Jin}}  \email{jinhai@tsinghua.edu.cn}
\author*[1]{\fnm{Wei} \sur{Cui}} \email{cui@tsinghua.edu.cn}

\affil[1]{\orgdiv{Department of Astronomy}, \orgname{Tsinghua University}, \orgaddress{\street{30 Shuangqing Road}, \city{Beijing}, \postcode{100084}, \country{the People's Republic of China}}}

\abstract{DIffuse X-ray Explorer (DIXE) is a proposed high-resolution spectroscopic survey mission onboard the China Space Station. Equipped with microcalorimeters based on the Transition-edge sensor technology, it aims to survey the hot gas in the Milky Way. The performance of DIXE depends on the understanding of non X-ray background (NXB), which can strongly affect observations of diffuse X-ray emission. In this work, we simulated the NXB of DIXE in a low-earth orbit (LEO) using \textsc{Geant4}. A detailed mass model of the payload was constructed, and the major sources of  NXB were identified, including cosmic rays, albedo neutrons and albedo photons. These components were implemented in \textsc{Geant4} with realistic angular and spectral distributions. We simulated the relevant physical processes of space radiation interacting with the instrument and calculated the resulting NXB. We also evaluated the delayed background from trapped protons in the South Atlantic Anomaly (SAA). Our simulations show that, at the geomagnetic equator and under solar minimum conditions, the NXB is on average $4.46 \times 10^{-2} ~\mathrm{counts~s^{-1}~cm^{-2}~keV^{-1}}$ in 0.1--10 keV energy band, with dominant contributions from the induced particles generated by primary cosmic protons. The NXB increases toward higher geomagnetic latitudes, reaching a maximum of $1.55 \times 10^{-1} ~\mathrm{counts~s^{-1}~cm^{-2}~keV^{-1}}$. The delayed background induced by the SAA decays rapidly after exiting the anomaly and becomes negligible within approximately 5 minutes. The simulated NXB is consistent with that of similar X-ray observatories in LEOs.}

\keywords{X-ray instrumentation, Non X-ray Background, Geant4, Low Earth orbit, DIXE}

\maketitle

%%%%%%%%%%%%%%%%%%%%%%%%%%%%%%%%%%%%%%%%%%%%%%%%%%%%%%%%%%%%%%
\section{Introduction}\label{sec: introduction}
Over the past few decades, there has been growing interest in the study of hot gas within the galactic ecosystem \citep{tumlinson_cgm_2017} (and references therein). The hot gas traces the feedback processes that play a crucial role in the evolution of galaxies, galaxy groups, and galaxy clusters, and is widely believed to account for the so-called `missing baryons' in the universe \citep{cen_where_1999, bregman_search_2007}. Such hot gas is best observed in the soft X-ray band. However, current X-ray observatories such as XMM-Newton and Chandra generally lack either the sensitivity to detect the faint diffuse emission or the spectral resolution to resolve diagnostic metal emission lines, which characterize the physical properties, including temperature, density and elemental abundances, of the hot gas. With the launch of XRISM \citep{xrismscienceteam_xrism_2020}, high-resolution X-ray spectroscopic observations are now possible and are expected to advance further with future missions such as HUBS \citep{cui_hubs_2020} and NewAthena \citep{cruise_newathena_2025}.

The Diffuse X-ray Explorer (DIXE; \cite{jin_dixe_2024}) is a proposed high-resolution X-ray spectroscopic survey experiment to be deployed onboard the China Space Station (CSS). During its planned three-year mission, DIXE will perform a sky survey while orbiting Earth at an altitude of approximately $400~\rm km$ in a low-earth orbit (LEO). The key scientific objective is to investigate the diffuse hot gas in the Milky Way, including the Local Hot Bubble, the eROSITA bubble and the Galactic hot halo. The DIXE payload consists of three major components \citep{jin_dixe_2024}: the scientific instruments, the supporting electronics, and the mechanical interface with CSS (see Fig. 2 in \cite{jin_dixe_2024}). The instrument employs a $10\times 10$ array of microcalorimeter detectors based on superconducting Transition-edge sensor (TES) technology. The detector array has an effective area of approximately $1 ~\rm cm^2$ and operates in the 0.1--10 keV energy band. The detectors are cooled to below 100 mK using a series of mechanical coolers and a two-stage Adiabatic Demagnetization Refrigerator (ADR), enabling an energy resolution of better than 6 eV at 6 keV. A collimator provides a $10\degr \times 10\degr$ field of view (FOV). As a first attempt to apply TES microcalorimeters to an astronomical application, DIXE will not carry anti-coincidence detectors (ACD) or focusing optics. Further details about the instrument design can be found in \cite{liu_DA_2024} and Section 4 of \cite{jin_dixe_2024}.

With the large FOV and high spectral resolution, DIXE can effectively probe the large structures of hot gas in the Milky Way and obtain their high-resolution spectra. However, the detection of such faint and diffuse sources is strongly limited by background noise. As such, an accurate assessment of the expected background levels is essential for evaluating the instrument’s observational capabilities. For space-based X-ray instruments, the background is dominated by the non X-ray background (NXB), which arises primarily from the interaction of energetic cosmic particles with the instrument. Cosmic particles contribute to the NXB through two principal mechanisms. First, they may traverse the instrument as minimum ionizing particles (MIPs; \cite{campana_bkg_2024}), depositing energy directly in the detector that is indistinguishable from genuine X-ray events. Second, they can interact with surrounding materials, producing secondary particles that subsequently deposit energy in the detector or trigger further interactions. In this work, we refer to the former particles and corresponding background as the `incident' particles and background and the latter ones as the `induced' particles and background. In some literature, the former is called `primary' particles and background, and the latter is called `secondary' particles and background. Here, the terminology of `incident' and `induced' is adopted to avoid confusion with the conventional classification of cosmic rays into primary and secondary particles, which is discussed in Section \ref{sec: space radiation environment}.

To model the physical processes responsible for the generation of the NXB, we employ the Monte Carlo simulation toolkit \textsc{Geant4}\footnote{\url{https://geant4.web.cern.ch/}} \citep{agostinelli_geant4_2003, allison_geant4_2006, allison_geant4_2016} version 11.4.0. \textsc{Geant4} is a widely used and highly versatile framework for simulating particle-matter interactions and has been extensively applied in high-energy physics, nuclear physics, space science, and medical research. This article is organized as follows. In Section \ref{sec: space radiation environment}, we describe the space radiation environment in LEOs and the space particle spectra adopted in the simulation. The simulation configuration, including the mass model and physics lists, is presented in Section \ref{sec: simulation configuration}. The results and their implications are discussed in Section \ref{sec: results and discussion}, followed by conclusions in Section \ref{sec: conclusion}.

%%%%%%%%%%%%%%%%%%%%%%%%%%%%%%%%%%%%%%%%%%%%%%%%%%%%%%%%%%%%%%
\section{Space radiation environment}\label{sec: space radiation environment}

The space radiation environment in LEO is primarily composed of cosmic rays, albedo neutrons, albedo photons, and energetic particles trapped in the South Atlantic Anomaly (SAA). In this section, we describe the characteristics and spectral models of these components as adopted in our simulations.

\subsection{Cosmic rays}\label{subsec: cosmic rays}

The cosmic rays in LEOs consist of both primary and secondary particles. Primary cosmic particles originate from galactic and extragalactic sources and are modulated by solar activity and the geomagnetic field before reaching near-Earth space. Secondary cosmic rays are produced by interactions between primary particles and the Earth’s atmosphere and typically exhibit lower energies than their primary counterparts.

Based on observations by the Alpha Magnetic Spectrometer (AMS) onboard the International Space Station (ISS) \citep{alcaraz_cosmic_2000, alcaraz_leptons_2000}, \cite{mizuno_cosmicray_2004} developed analytical models for the spectra of both primary and secondary cosmic rays in LEOs. In the following, we summarize the formulation of these models as implemented in this work; further details can be found in the original publication.

\subsubsection{Primary cosmic particles}\label{subsubsec: primary cosmic particles}

The unmodulated spectrum of cosmic rays entering the solar system can be described by a power-law function of rigidity \citep{mizuno_cosmicray_2004}:
\begin{equation}
    F_{\mathrm{unmod}}(E_\mathrm{k}) = A \left[ \frac{R(E_\mathrm{k})}{\mathrm{GV}} \right]^{-a}
\end{equation}
Here, $E_\mathrm{k}$ is the kinetic energy of the particle, and $R$ is the rigidity, defined as a function of momentum $p$ and charge $q$:
\begin{equation}
    R = \frac{pc}{q} = \frac{1}{Z\mathrm{e}} \sqrt{E_\mathrm{k}^2+2Mc^2E_\mathrm{k}}
\end{equation}
with $Z$ and $M$ denoting the particle charge and mass, respectively. 

Solar wind can decelerate the cosmic particles, thus reducing the flux of low-energy cosmic particles. The solar-modulated spectrum is given by \citep{mizuno_cosmicray_2004}: 
\begin{equation}
    F_\mathrm{mod}(E_\mathrm{k}) = F_\mathrm{unmod}(E_\mathrm{k}+Z\mathrm{e}\phi) \times \frac{\left(E_\mathrm{k}+Mc^2\right)^2-\left(Mc^2\right)^2}{\left(E_\mathrm{k}+Mc^2+Z\mathrm{e}\phi\right)^2-\left(Mc^2\right)^2}
\end{equation}
where $\phi$ is the solar modulation potential, typically ranging from $0.55~\rm GV$ (solar minimum) to $1.1~\rm GV$ (solar maximum).

Near Earth, the geomagnetic field can stop cosmic particles from entering Earth's atmosphere, especially at low geomagnetic latitudes. This effect manifests itself in the spectrum as a suppression in the low-rigidity (low-energy) region. The combined effect of solar and geomagnetic modulation \citep{mizuno_cosmicray_2004} is expressed as:
\begin{equation}
    F_\mathrm{primary}(E_\mathrm{k}) = F_\mathrm{mod}(E_\mathrm{k}) \times \frac1{1+\left[R(E_\mathrm{k}) / R_{\mathrm{cut}}\right]^{-r}}
\end{equation}
The term $\frac1{1+\left[R(E_\mathrm{k}) / R_{\mathrm{cut}}\right]^{-r}}$ is the geomagnetic cut-off function, where the parameter $r$ characterizes the sharpness of the flux transition around the cut-off rigidity (COR) $R_{\mathrm{cut}}$. Following \cite{mizuno_cosmicray_2004}, we adopt $r=6$ for protons and alpha particles, and $r=12$ for electrons and positrons.

The cut-off rigidity $R_{\mathrm{cut}}$ is approximated using the simplified Störmer equation \citep{smart_rcut_2005}:
\begin{equation}
    R_{\mathrm{cut}} = 14.9 \times \left(1+\frac h{R_{\oplus}}\right)^{-2.0} \left(\cos\theta_\mathrm{M}\right)^4 ~\mathrm{GV}
    \label{Rcut}
\end{equation}
where $h$ is the orbit altitude, $R_{\oplus} = 6371 ~\rm km$ is the Earth's radius , and $\theta_\mathrm{M}$ is the geomagnetic latitude.

For the CSS orbit with an inclination of $42\degr$ and considering the $\sim 11\degr$ tilt between the geographic and geomagnetic axes, we approximate the maximum geomagnetic latitude that CSS can reach as $53\degr$.  

Combining the above effects, the primary cosmic-ray spectrum in LEOs is given by
\begin{align}
    F_\mathrm{primary}(E_\mathrm{k}) = & A \left[\frac{R(E_\mathrm{k}+Ze\phi)}{\mathrm{GV}}\right]^{-a} \times\frac{\left(E_\mathrm{k}+Mc^2\right)^2-\left(Mc^2\right)^2}{\left(E_\mathrm{k}+Mc^2+Z\mathrm{e}\phi\right)^2-\left(Mc^2\right)^2} \nonumber \\ 
    &\times\frac1{1+\left[R(E_\mathrm{k})/R_{\mathrm{cut}}\right]^{-r}}
\end{align}

\begin{figure}[t!]
    \centering
    \includegraphics[width=\linewidth]{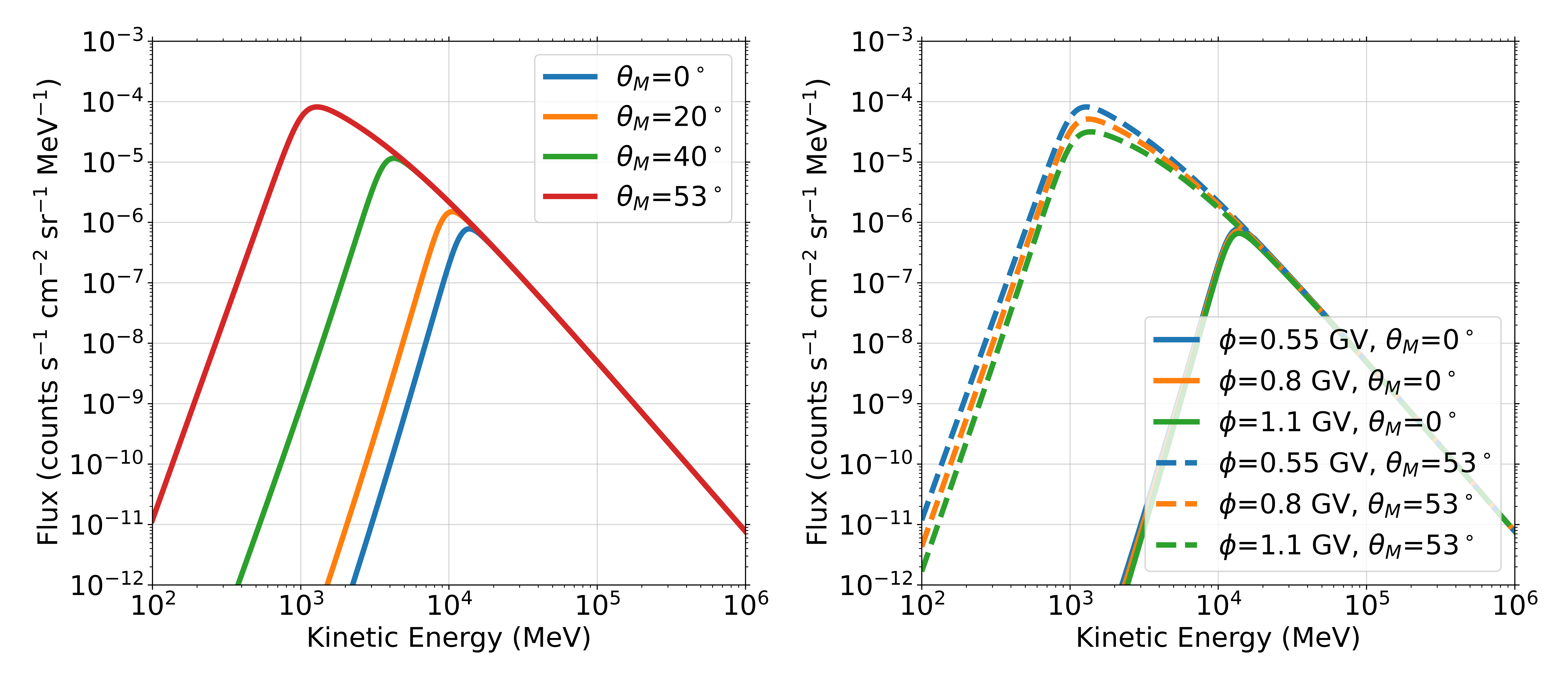}
    \caption{Spectrum of primary cosmic protons for $h=400 ~\rm km$. Left: spectrum for $\phi=0.55 ~\rm GV$ at different geomagnetic latitudes $\theta_\mathrm{M}$. Right: spectrum for different solar activities. The solid lines, which nearly overlap with each other, show the flux at $\theta_\mathrm{M}=0\degr$ and different different $\phi$, from solar minimum ($\phi = 0.55 ~\rm GV$) to solar maximum ($\phi=1.1 ~\rm GV$). The dashed lines show the flux at $\theta_\mathrm{M}=53.5\degr$ and different $\phi$}
    \label{fig_p_p_spectrum}
\end{figure}

The normalization $A$ and spectral index $a$ are taken from Section. 3.2--3.4 of \cite{mizuno_cosmicray_2004}. The left panel of Fig. \ref{fig_p_p_spectrum} shows the dependence of the primary proton spectrum on geomagnetic latitude. As $\theta_\mathrm{M}$ increases, the peak of the differential spectrum shifts toward lower energy and higher flux. This indicates that the geomagnetic field can effectively `cut off' low-energy cosmic particles and this `cut-off' effect is reduced at high latitudes. At low geomagnetic latitudes, the flux varies only weakly with latitude, whereas at high latitudes it increases rapidly, consistent with the $\cos\theta_\mathrm{M}$ dependence in Equation \eqref{Rcut}. This relation is also demonstrated in Fig. \ref{fig_bkg_latitude}. The influence of solar activity is illustrated in the right panel of Fig. \ref{fig_p_p_spectrum}. While stronger solar activity reduces the overall flux, its effect is modest compared to geomagnetic modulation, particularly at low geomagnetic latitudes. Similar trends apply to other primary cosmic-ray species.

\subsubsection{Secondary cosmic particles}\label{subsubsec: secondary cosmic particles}

Secondary cosmic rays are produced by interactions between primary cosmic rays and the Earth’s atmosphere. Their spectra depend on geomagnetic latitude and are described by piecewise power-law models \citep{mizuno_cosmicray_2004}.

For secondary protons with energies above 100~MeV, the spectrum is modeled either as a broken power law \citep{mizuno_cosmicray_2004}:
\begin{align}
    &F_0\left(\frac{E_\mathrm{k}}{100\mathrm{~MeV}}\right)^{-a},\quad100\mathrm{~MeV}\leq E_\mathrm{k}\leq E_{\mathrm{bk}}, \nonumber \\ 
    &F_0\left(\frac{E_{bk}}{100\mathrm{~MeV}}\right)^{-a}\left(\frac{E_\mathrm{k}}{E_{\mathrm{bk}}}\right)^{-b},\quad E_\mathrm{bk}\leq E_\mathrm{k}
    \label{brokenPL}
\end{align}
 or as a cut-off power law: 
\begin{equation}
    F_1\left(\frac{E_\mathrm{k}}{\mathrm{GeV}}\right)^{-a}\exp-\left(\frac{E_\mathrm{k}}{E_{\mathrm{cut}}}\right)^{-a+1},\quad100\mathrm{~MeV}\leq E_\mathrm{k}
\end{equation}
Below 100 MeV, where AMS data are unavailable, a simple power law with an index of $-1$ is adopted in \cite{mizuno_cosmicray_2004}: 
\begin{equation}
    F_0\left(\frac{E_\mathrm{k}}{100~\mathrm{ MeV}}\right)^{-1},\quad1~\mathrm{ MeV}\leq E_\mathrm{k}\leq100~\mathrm{ MeV}
    \label{PL-1}
\end{equation}

Secondary electrons and positrons share the same spectral shape but differ in normalization, determined by the $e^+ / e^-$ ratio \citep{Golden_positron_1994}. Their spectra above 100 MeV are described either by a power law:
\begin{equation}
    F_{0}{\left(\frac{E_\mathrm{k}}{100\mathrm{~MeV}}\right)}^{-a},\quad100 ~\mathrm{MeV}\leq E_\mathrm{k}
\end{equation}
or a broken power law as in Equation \eqref{brokenPL}, or a power law with a hump \citep{mizuno_cosmicray_2004}:
\begin{equation}
    F_{0} {\left(\frac{E_\mathrm{bk}}{100\mathrm{~MeV}}\right)}^{-a} + F_{1} {\left(\frac{E_\mathrm{k}}{\mathrm{GeV}}\right)}^{b}\exp{-\left(\frac{E_\mathrm{k}}{E_{c}}\right)}^{b+1},\quad 100\mathrm{~MeV}\leq E_\mathrm{k}
\end{equation}
Below 100 MeV, the same power law with an index of $-1$ as Equation \eqref{PL-1}, is used \citep{mizuno_cosmicray_2004}. 

\begin{figure}[t!]
    \centering
    \includegraphics[width=\textwidth]{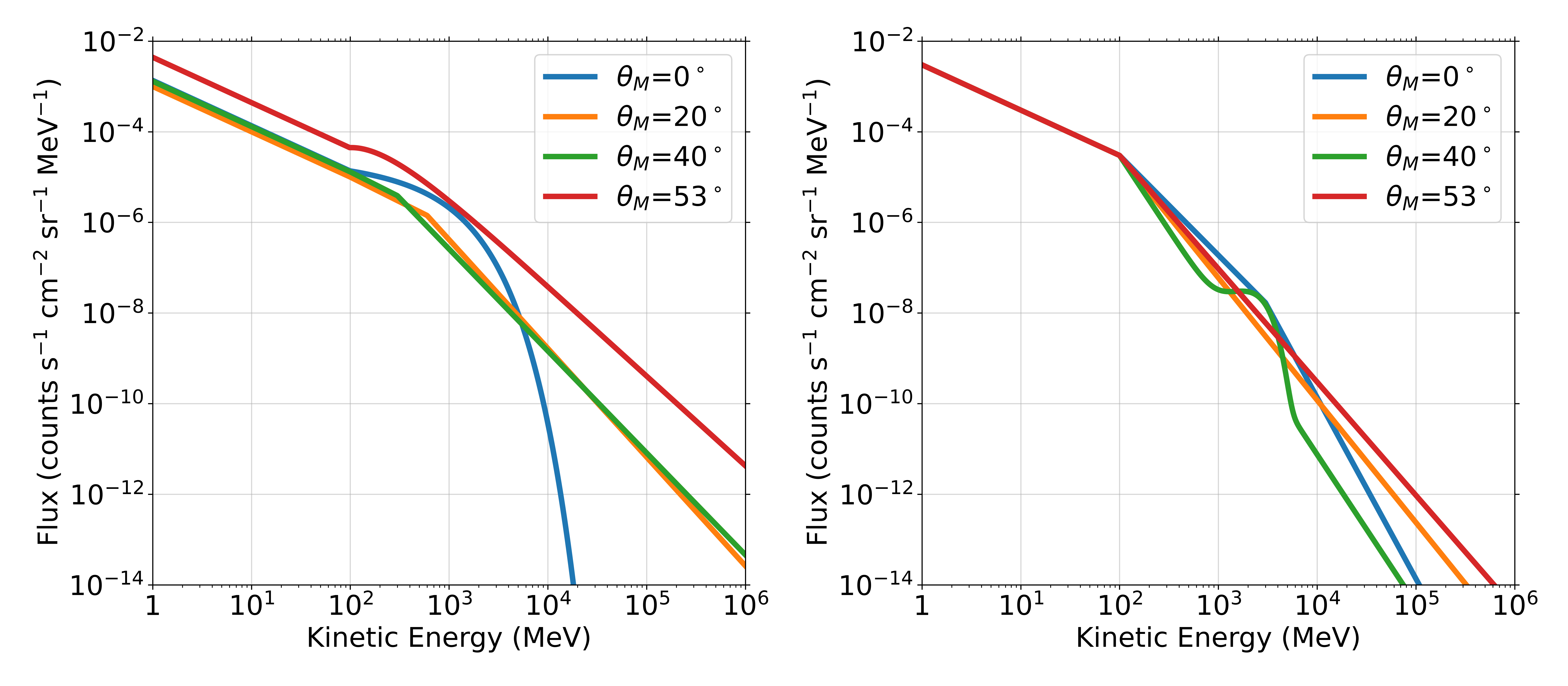}
    \caption{Spectrum of downward secondary cosmic protons (left) and electrons (right) for $h=400 ~\rm km$ and $\phi =0.55 ~\rm GV$ at different $\theta_\mathrm{M}$}
    \label{fig_s_spectrum_latitude}
\end{figure}

The model parameters are taken from Tables 1 and 2 of \cite{mizuno_cosmicray_2004}. We show the spectrum of downward secondary protons and electrons at different geomagnetic latitudes as an example in Fig. \ref{fig_s_spectrum_latitude}. Compared to primary cosmic rays, the flux of secondary particles exhibits a weaker and non-monotonic dependence on geomagnetic latitude, decreasing at intermediate latitudes before increasing again toward higher latitudes, as also shown in Fig. \ref{fig_bkg_latitude}.

\subsection{Albedo Neutrons}\label{subsec: albedo neutrons}

Interactions between cosmic rays and the Earth’s atmosphere also produce neutrons, a fraction of which can propagate upward into LEOs and contribute to the instrumental background. Current models of albedo neutrons are mainly based on Monte Carlo simulations, including those by \cite{armstrong_neutron_1973, kole_neutron_2015}, and the Models for Atmospheric Ionizing Radiation Effects (MAIRE\footnote{\url{http://www.radmod.co.uk/maire}}).

In this work, we adopt the neutron spectrum from \cite{kole_neutron_2015}. While the neutron flux consists of upward and downward components, the downward flux becomes negligible at LEO altitudes. Therefore, we consider only the upward (albedo) neutrons. 

The spectrum is divided into four energy intervals: $8 ~\mathrm{keV} - 0.9 ~\mathrm{MeV}$, $0.9  - 15 ~\mathrm{MeV}$, $15 - 70 ~\mathrm{MeV}$, and $70 - 1000 ~\mathrm{MeV}$, each with a power law:
\begin{align}
    & F = A \left( \frac{E}{\mathrm{MeV}} \right)^{-\alpha} \nonumber \\
    & F = B \left( \frac{E}{\mathrm{MeV}} \right)^{-\beta} \nonumber \\
    & F = C \left( \frac{E}{\mathrm{MeV}} \right)^{-\gamma} \nonumber \\
    & F = D \left( \frac{E}{\mathrm{MeV}} \right)^{-\delta}
    \label{neutron}
\end{align}
where $F$ is the neutron flux in units of $\rm counts ~cm^{-2} ~s^{-1} ~MeV^{-1}$, and $A$, $B$, $C$,$D$ are the normalizations of each segment. 

The normalization factor $A$ in the first energy interval depends on atmospheric pressure (height) $h$ (hPa), modeled as:
\begin{equation}
    A = (ah+b)e^{-h/c} + d
    \label{A}
\end{equation}
As can be seen in Fig. 3 of \cite{kole_neutron_2015}, $A$ nearly stays constant for $h < 10^{-2} ~\mathrm{hPa}$. At LEO altitudes, where the atmospheric pressure is well below $10^{-8} ~\mathrm{hPa}$, the normalization factor $A$ becomes effectively independent of altitude and can be evaluated by setting $h=0$ in Equation \eqref{A}, yielding
\begin{equation}
    A = b + d
    \label{A_simple}
\end{equation}

The coefficients $b$ and $d$ in Equation \eqref{A_simple} depend on both geomagnetic latitude $\theta_\mathrm{M}$ and the solar activity parameter $S = (\phi - 0.25 ~\mathrm{GV}) / 0.859 ~\mathrm{GV}$ \citep{cumani_background_2019}, where $\phi$ is the solar modulation potential introduced in Section \ref{subsec: cosmic rays}. The coefficients are expressed as:
\begin{align}
    & \mathrm{b}=1.4 \times 10^{-2}+(1.4-0.9 S) \times 10^{-1}\left[1-\tanh (180-3.5 \theta_\mathrm{M})\right] \nonumber \\
    & \mathrm{d}=-8.0 \times 10^{-3}+(6.0-1.05) \times 10^{-3}\left[1-\tanh (180-4.4 \theta_\mathrm{M})\right]
\end{align}

The spectral indices in Equation \eqref{neutron} are also functions of atmospheric pressure (height) $h$, and hence can be considered constant at LEO altitudes:
\begin{align}
    & \alpha=-0.290 e^{-h / 7.5} + 0.735  = 0.445  \nonumber \\
    & \beta=-0.247 e^{-h / 36.5} + 1.40  = 1.153 \nonumber \\
    & \gamma=-0.40 e^{-h / 40} + 0.90  = 0.5 \nonumber \\
    & \delta=-0.46 e^{-h / 100} + 2.53 = 2.07
\end{align}

To ensure continuity across energy intervals, the normalization constants $B$, $C$ and $D$ are calculated from $A$ and the corresponding spectral indices: 
\begin{align}
    & B = A \cdot 0.9^{-\alpha+\beta} \nonumber \\
    & C = B \cdot 15^{-\beta+\gamma} \nonumber \\
    & D = C \cdot 70^{-\gamma+\delta} 
\end{align}

\begin{figure}[t!]
    \centering
    \includegraphics[width=\textwidth]{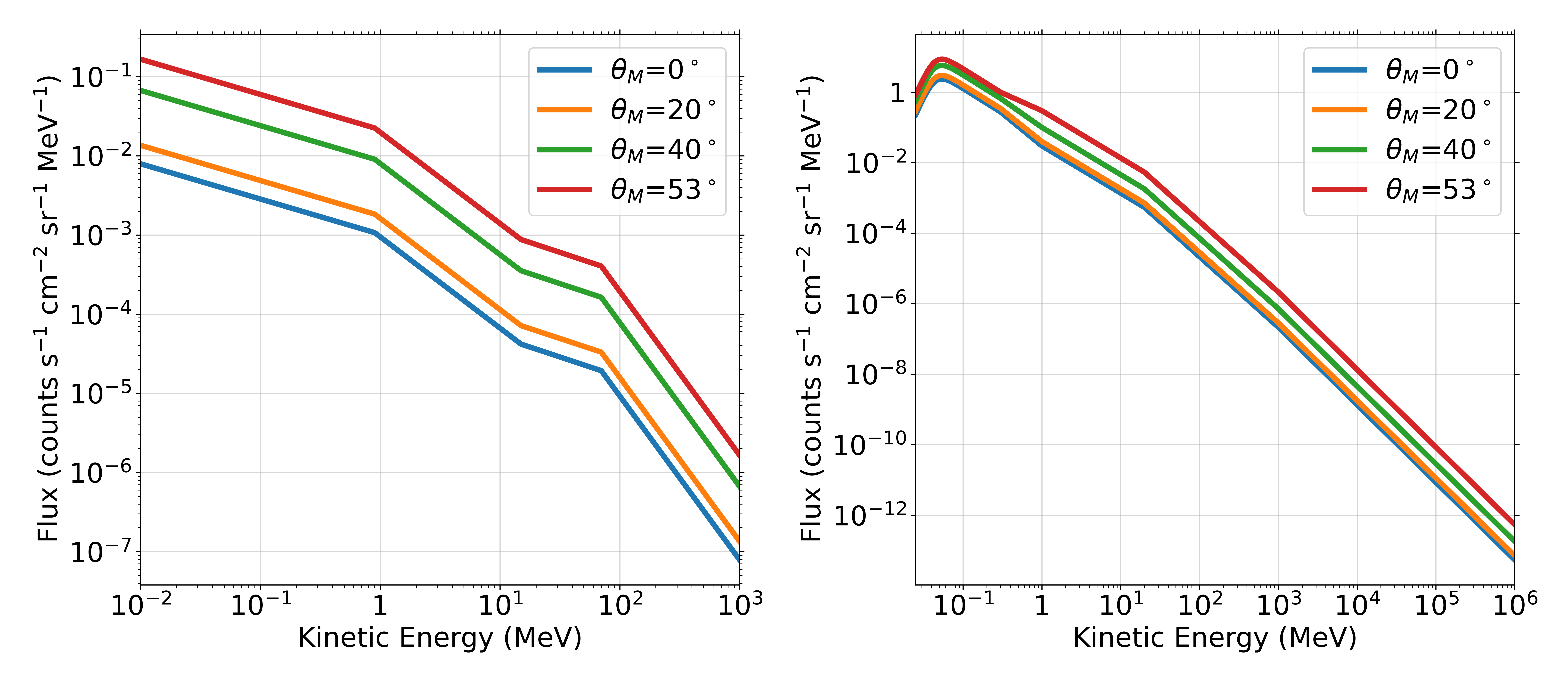}
    \caption{Spectrum of albedo neutrons (left) and albedo photons (right) for $h=400 \rm ~km$ and $\phi=0.55 ~\rm GV$ at different geomagnetic latitudes $\theta_\mathrm{M}$}
    \label{fig_al_spectrum_latitude}
\end{figure}

Finally, we divide the flux in Equation \eqref{neutron} by $2\pi$ solid angle to obtain the differential flux per unit solid angle. The resulting albedo neutron spectra at different geomagnetic latitudes are shown in the left panel of Fig. \ref{fig_al_spectrum_latitude}, illustrating a strong dependence on geomagnetic latitude.

\subsection{Albedo photons}

The albedo photons are X-ray and $\gamma$-ray photons emitted upward from the Earth's atmosphere, produced by interactions between incident cosmic particles and atmospheric constituents. For the energy range of 25–-300 keV, we adopt the model derived from Monte Carlo simulations presented in \cite{sazonov_gamma_2007}. This model yields a spectrum peaking at $\sim 50 ~\mathrm{keV}$, which can be well approximated by
\begin{align}
    F(E_\mathrm{k})=\frac{C}{(E_\mathrm{k} / 44 \mathrm{keV})^{-5}+(E_\mathrm{k} / 44 ~\mathrm{keV})^{1.4}}
\end{align}
The normalization factor $C$, in units of $\rm counts ~s^{-1} ~cm^{-1} ~sr^{-1} ~{MeV}^{-1}$, depends on the solar modulation factor $\phi$, geomagnetic cut-off rigidity $R_{\mathrm{cut}}$, and the zenith angle $\theta$ ($\mu=\cos \theta$). It is given by: 
\begin{align}
    C=\frac{3 \mu(1+\mu)}{5 \pi} \frac{1.47 \times 17.8\left[(\phi / 2.8)^{0.4}+(\phi / 2.8)^{1.5}\right]^{-1}}{\left[1+\left\{R_{\mathrm{cut}} /\left[1.3 \phi^{0.25}\left(1+2.5 \phi^{0.4}\right)\right]\right\}^2\right]^{0.5}}
    \label{C_gamma}
\end{align}

In \cite{sazonov_gamma_2007}, the original Störmer equation is used to calculate $R_{\mathrm{cut}}$ (see Equation (8) therein). However, in our case, DIXE is assumed to always point towards the zenith, which allows us to simplify the calculation of $R_{\mathrm{cut}}$ using Equation \eqref{Rcut} in Section \ref{subsubsec: primary cosmic particles}. The parameter $\mu$ in Equation \eqref{C_gamma} also remains constant at zero due to the fixed zenith angle.

For albedo photons with energy above 1 MeV, we employ the model from \cite{mizuno_cosmicray_2004}, which is based on satellite and balloon flight data \citep{thompson_gamma_1974, imhof_gamma_1976, gurian_gamma_1979, ryan_gamma_1979}. The spectrum, expressed in units of $\rm counts ~s^{-1} ~cm^{-1} ~sr^{-1} ~{MeV}^{-1}$, is described by a piecewise power-law function:
\begin{align}
    F(E_\mathrm{k})= 
    \begin{cases}
    0.101 E^{-1.34} & 1 \leq E < 20 ~\mathrm{MeV} \\ 
    0.729 E^{-2.0} & 20 \leq E < 10^3 ~\mathrm{MeV} \\ 
    2.9  E^{-2.2} & 10^3 \leq E < 10^6 ~\mathrm{MeV}
    \end{cases}
\end{align}
This spectrum is generated from data measured at $R_{\mathrm{cut}}=4.5 ~\rm GV$. The flux of albedo photons above 80 MeV is anti-correlated with the cut-off rigidity, following a dependence of $R_{\mathrm{cut}}^{-1.13}$ \citep{mizuno_cosmicray_2004}. Accordingly, we apply a scaling factor of $(R_{\mathrm{cut}}/4.5~\rm GV)^{-1.13}$ to account for different geomagnetic latitudes.

To bridge the energy gap between 0.3 MeV and 1 MeV, we interpolate using a power-law function fitted to the fluxes at 0.3 MeV and 1 MeV. The resulting spectra at different geomagnetic latitudes are shown in the right panel of Fig. \ref{fig_al_spectrum_latitude}. As with other space radiation components, the dependence on geomagnetic latitude is weak at low latitudes but becomes increasingly significant at higher latitudes, particularly for photon energies above 300 keV.

\begin{figure}[t!]
    \centering
    \includegraphics[width=0.6\textwidth]{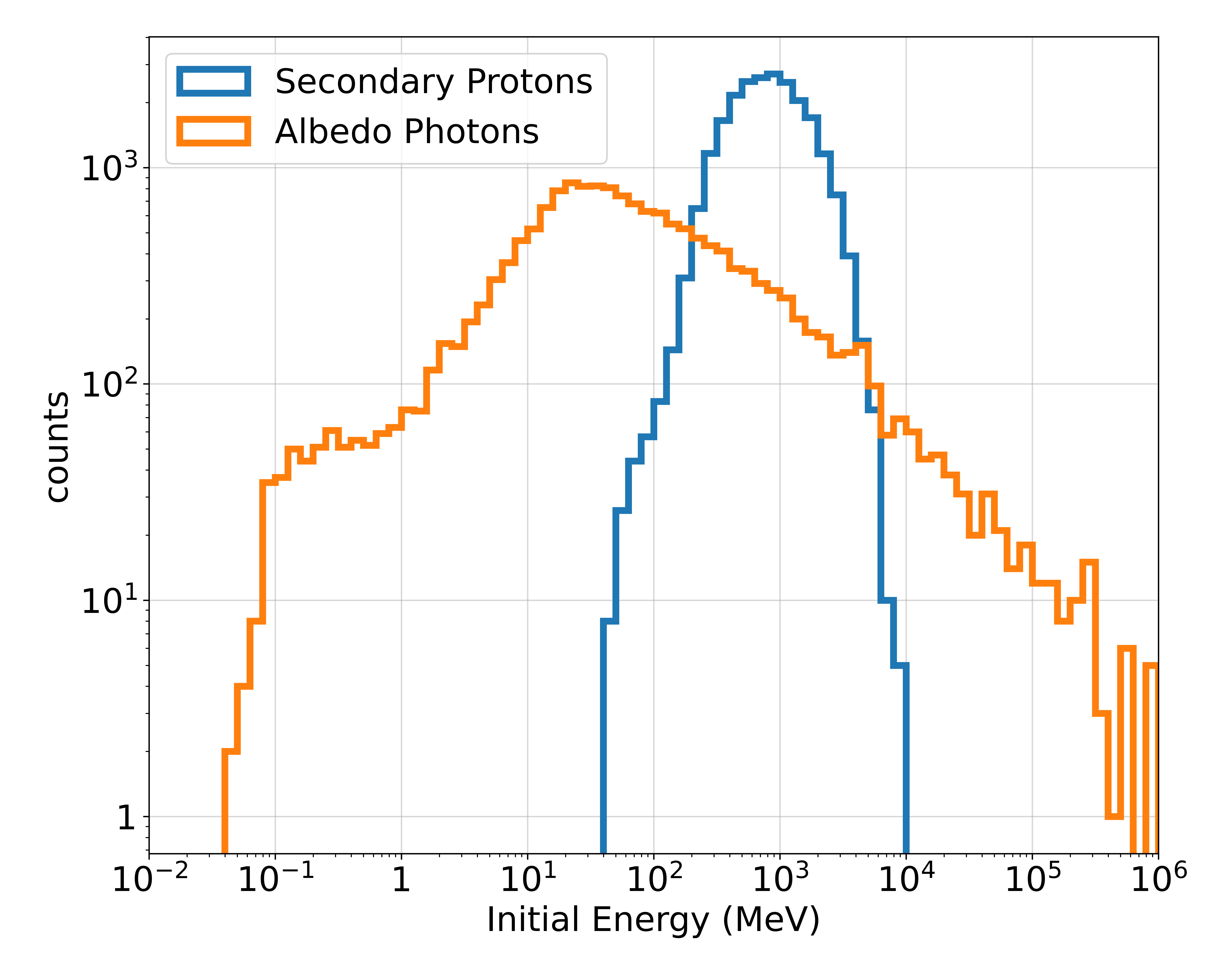}
    \caption{Initial energy distribution of secondary cosmic protons (blue histogram) and albedo photons (orange histogram) that generates background events in the detectors}
    \label{fig_initial_energy}
\end{figure}

In some previous NXB simulations \citep{campana_loft_2013, galgoczi_configuration_2021} for satellites in LEOs, only albedo photons with energies below 20 MeV have been considered. To assess the importance of higher-energy albedo photons, we examined the initial energies of albedo photons that generate background events in the detectors. As shown in Fig. \ref{fig_initial_energy}, albedo photons with initial energies higher than 20 MeV provide more than half of the background. This demonstrates that albedo photons above 20 MeV must also be included in NXB simulations.  

\begin{figure}[t!]
    \centering
    \includegraphics[width=0.9\textwidth]{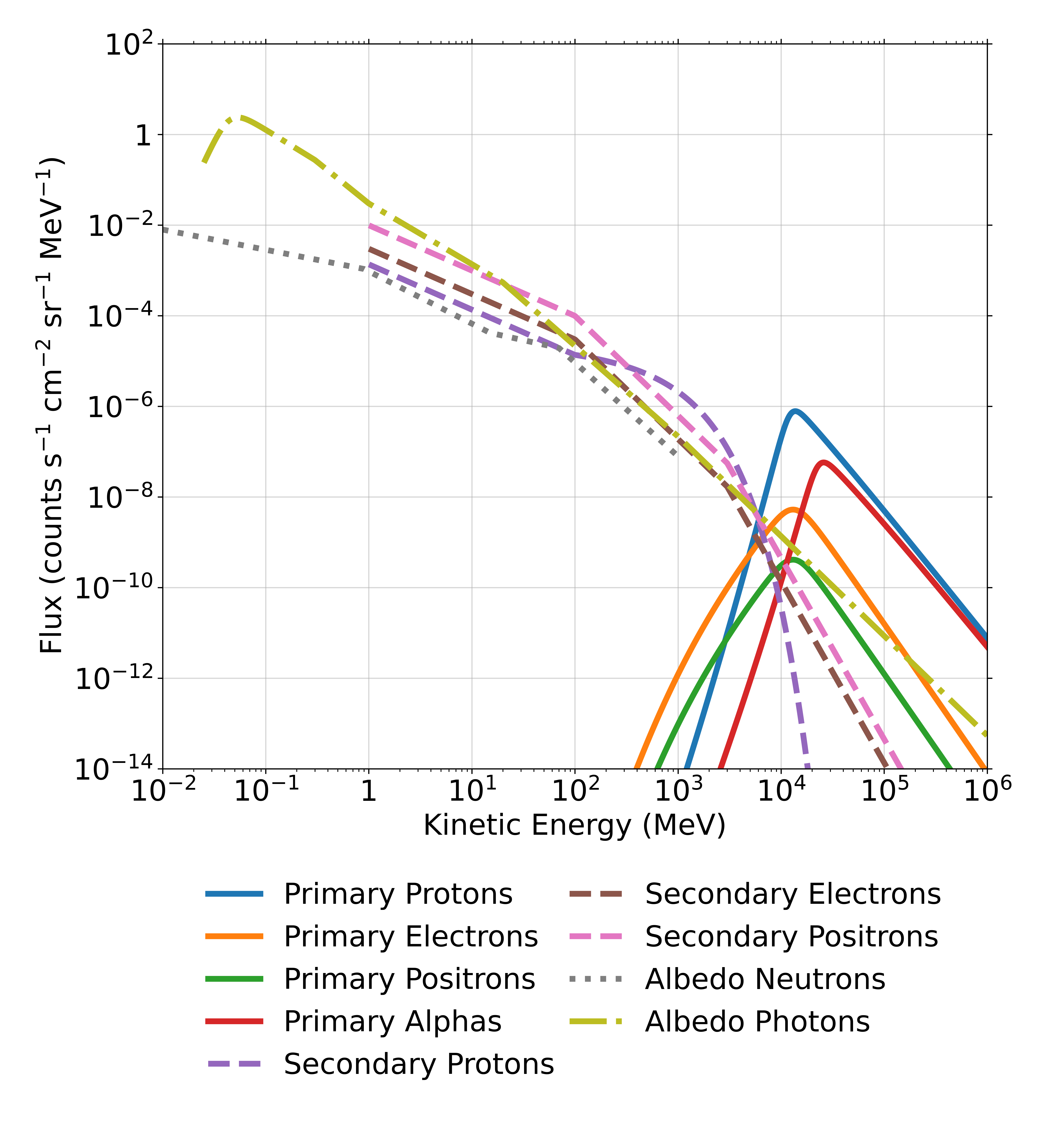}
    \caption{Total spectrum of all space radiation in LEOs at $\theta_\mathrm{M}=0\degr$, $\phi=0.55 ~\rm GV$ and $h=400 ~\rm km$. Solid lines show primary protons (blue), electrons (orange), positrons (green), and alpha particles (red). Dashed lines show secondary protons (purple), electrons (brown), and positrons (pink). The dotted gray line indicates albedo neutrons, and the dash-dotted yellow line represents albedo photons}
    \label{fig_spectrum_all}
\end{figure}

Fig. \ref{fig_spectrum_all} summarizes the spectra of all space radiation components considered in this work. Secondary cosmic particles, albedo neutrons, and albedo photons dominate the low-energy regime, while primary cosmic particles dominate at higher energies.

\subsection{SAA particles}
At altitudes between roughly 1000 km and 60000 km, the Earth is surrounded by the Van Allen radiation belts, where large populations of energetic charged particles are trapped. These radiation belts are formed through interactions between the solar wind, cosmic rays, and the Earth’s geomagnetic field \citep{campana_bkg_2024}. LEO lies below the main body of the radiation belts. However, the Van Allen radiation belts are symmetric about the Earth's magnetic axis, which is tilted with respect to the rotation axis, and the magnetic dipole center is offset from the Earth's geometric center. As a result, in a region over the South Atlantic Ocean, the inner radiation belt approaches closer to the Earth's surface (see Fig. 4 of \cite{campana_bkg_2024}), reaching altitudes as low as $\sim 200 ~\mathrm{km}$. This region is known as the South Atlantic Anomaly (SAA). 

Satellites traversing the SAA are exposed to intense fluxes of energetic charged particles. To mitigate potential damage or contamination, scientific operations are typically suspended during SAA passages. Nevertheless, the high particle fluxes can activate satellite materials, producing radioactive isotopes with half-lives long enough to emit radiation well after the spacecraft has exited the SAA. This radiation constitutes the so-called `delayed background', in contrast to the prompt background generated immediately by particle interactions

\begin{figure}[t!]
    \centering
    \includegraphics[width=\textwidth]{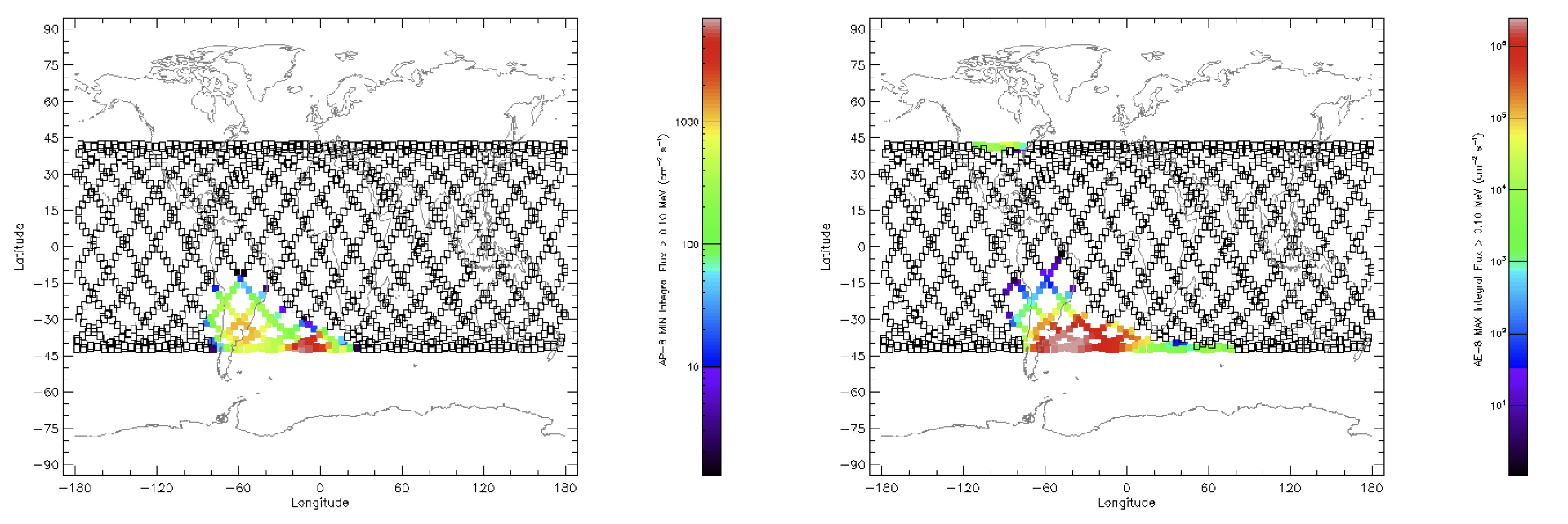}
    \caption{CSS orbit over one day and flux maps of trapped protons (left) and electrons (right) with energies $>0.1$ MeV at solar minimum, generated using SPENVIS}
    \label{fig_SAA_map}
\end{figure}

Fig. \ref{fig_SAA_map} shows the orbit of the CSS over one day, along with modeled proton and electron flux maps. These results, as well as the SAA proton and electron spectra discussed below, were obtained from SPENVIS\footnote{\url{https://www.spenvis.oma.be/}} using the AP8 \citep{sawyer_ap8_1976} and AE8 \citep{vette_ae8_1991} models. Proton and electron fluxes are remarkably higher within the SAA than elsewhere. 

\begin{figure}[t!]
    \centering
    \includegraphics[width=\textwidth]{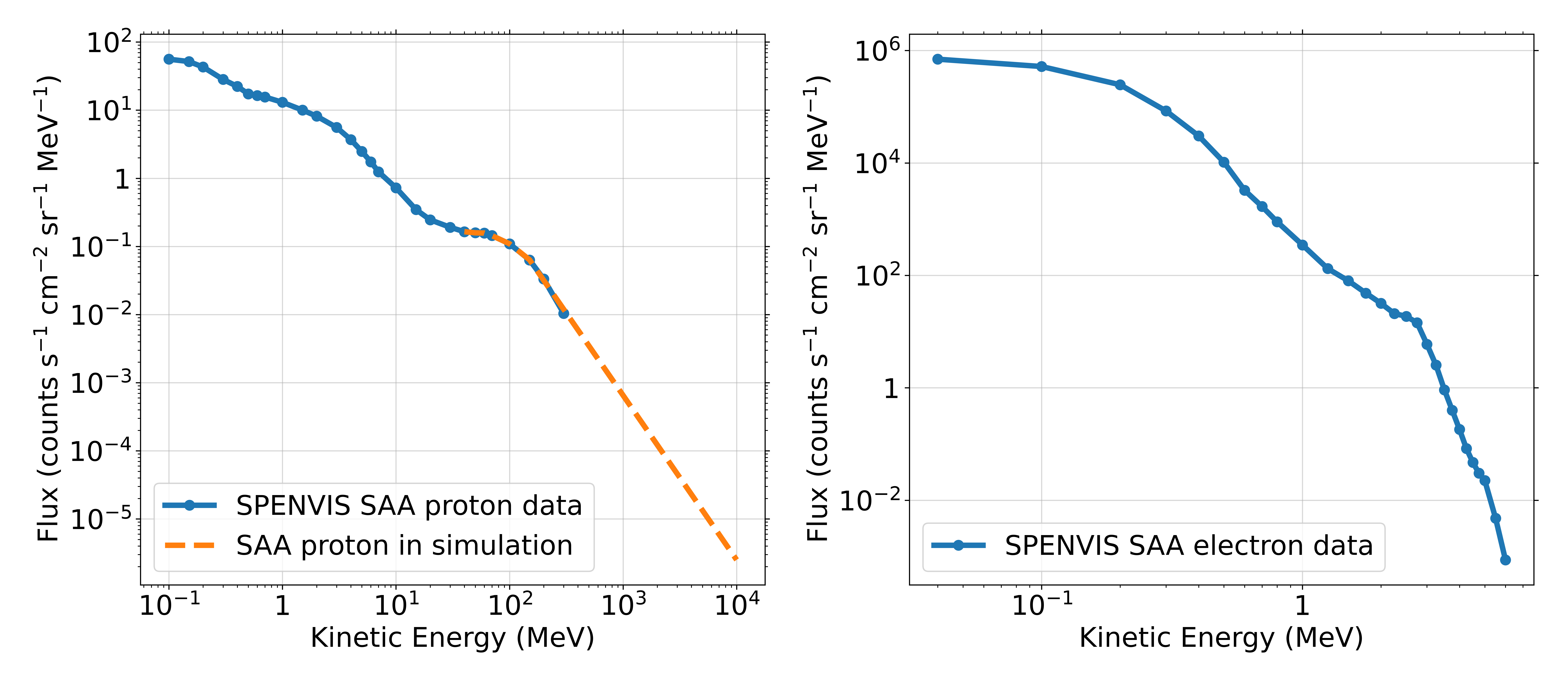}
    \caption{Spectra of trapped protons and electrons inside the SAA. Left: proton spectrum from SPENVIS (blue line) and the extrapolated power law used in simulations (orange dashed line), as given by Equation \eqref{saa_PL}. Right: trapped electron spectrum}
    \label{fig_SAA_spectrum}
\end{figure}

Fig. \ref{fig_SAA_spectrum} presents the trapped proton and electron spectra within the SAA. The SPENVIS output represents the average particle spectra over one day. Since the CSS spends approximately 8\% of its 1.5-hour orbital period inside the SAA \citep{feng_polar_2024}, the spectra are divided by 0.08 to obtain the corresponding in-situ fluxes. The fluxes are further divided by $4\pi$ sr to ensure consistency with the units used for other space radiation components.

The spectrum of trapped protons ends at 400 MeV. However, as shown in Fig. \ref{fig_initial_energy}, most background events generated by secondary cosmic protons arises from protons with initial energies above 100 MeV. The spectral shape of trapped protons in the SAA (left panel of Fig. \ref{fig_SAA_spectrum}) closely resembles that of secondary cosmic protons (see the left panel of Fig. \ref{fig_s_spectrum_latitude}), both following an approximate power-law behavior. To account for the contribution of higher-energy trapped protons, we fit the data points at 150, 200, and 300 MeV with a power law and extrapolate to $10^4 ~\rm MeV$. The fitted result is (in units of $\rm counts ~s^{-1} ~cm^{-1} ~sr^{-1} ~{MeV}^{-1}$)
\begin{align}
    F(E_\mathrm{k}) = 1.144\times10^4 E_\mathrm{k}^{-2.41}, \quad 150 \leq E_\mathrm{k} \leq 10^4 ~\rm MeV  
    \label{saa_PL}
\end{align}
In the simulation,  we use the SPENVIS data points for $E_\mathrm{k} \geq 40$ MeV together with the extrapolated spectrum, shown as the orange dashed line in the left panel Fig. \ref{fig_SAA_spectrum}.

The trapped electron spectrum (right panel of Fig. \ref{fig_SAA_spectrum}) is confined to lower energies than that of protons. Extrapolation to higher energies yields fluxes much lower than those of secondary cosmic electrons. Even the prompt background from trapped electrons will be lower than that from the cosmic particles. Therefore, the delayed background from trapped SAA electrons is not included in our simulations.

%%%%%%%%%%%%%%%%%%%%%%%%%%%%%%%%%%%%%%%%%%%%%%%%%%%%%%%%%%%%%%
\section{Simulation configuration}\label{sec: simulation configuration}

\subsection{Mass model}\label{subsec: mass model}

\begin{figure}[t!]
    \centering
    \includegraphics[width=0.8\textwidth]{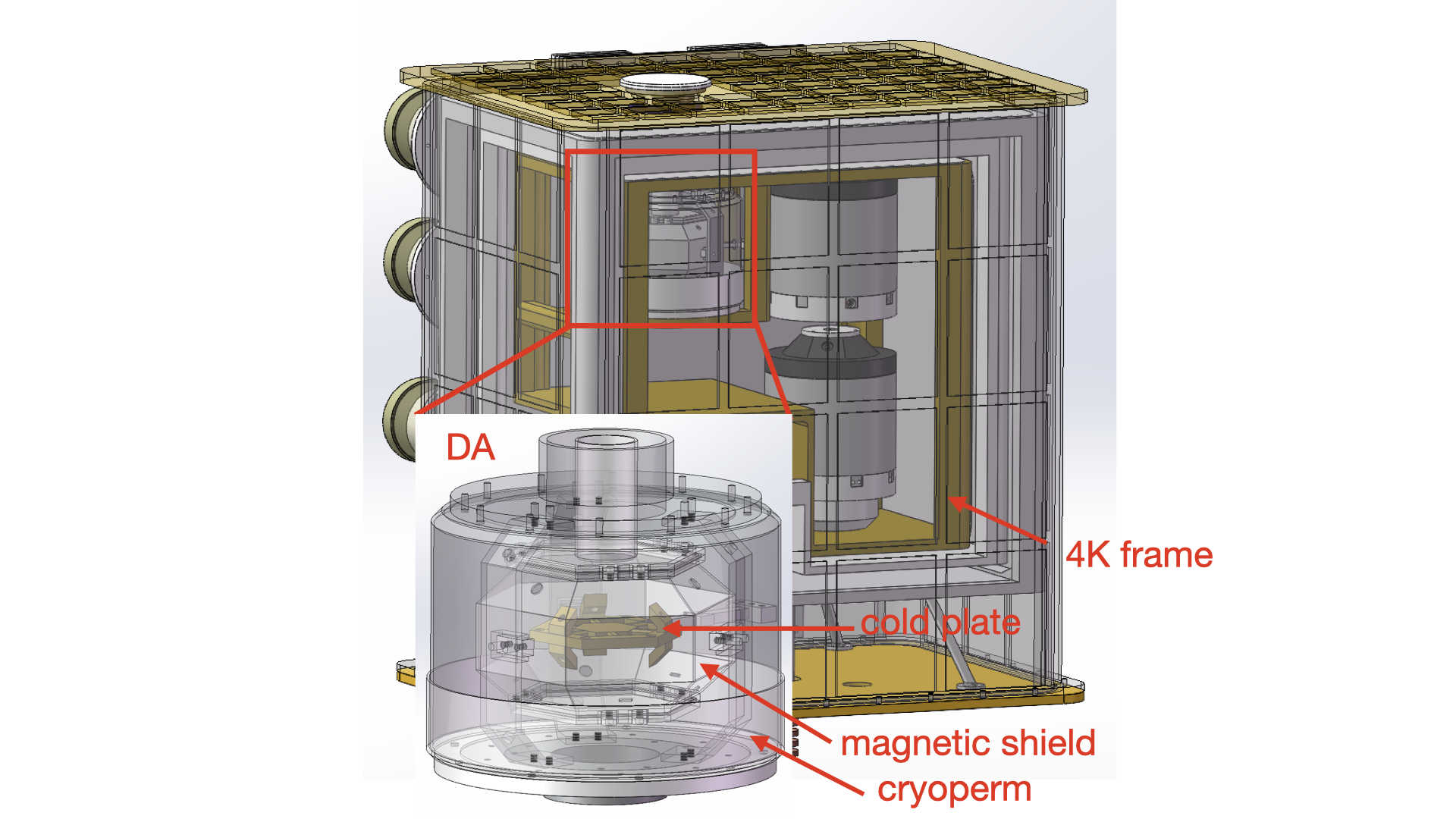}
    \caption{CAD model of DIXE (detector array not included) used to construct the mass model in the simulation. The zoomed-in view shows the DA. The positions of the cold plate, Nb magnetic shield, cryoperm and the 4K frame are labeled}
    \label{fig_mass_model}
\end{figure}

We constructed the mass model of DIXE used in the simulations based on the current payload instrument design (see Fig. \ref{fig_mass_model} and Fig. 3 of \cite{jin_dixe_2024}). Supporting electronics and the space station adapters were not included, as their designs have not yet been finalized.

\begin{figure}[t!]
    \centering
    \includegraphics[width=\textwidth]{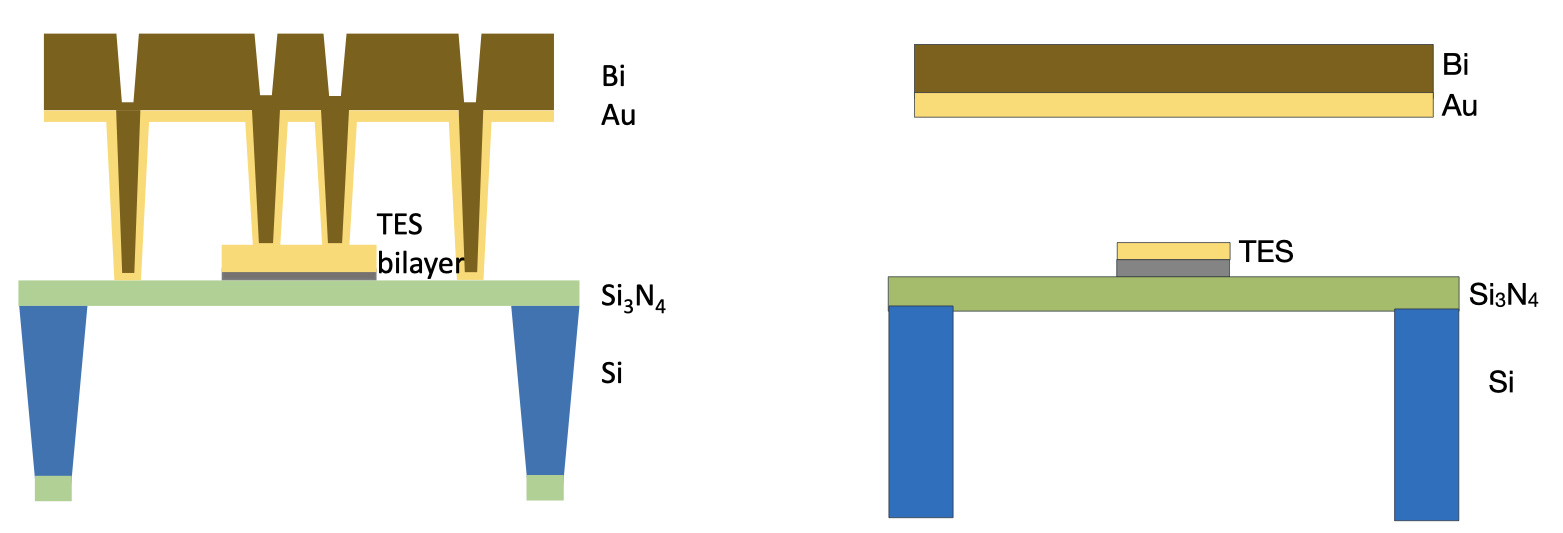}
    \caption{Left: the original (not to scale) microcalorimeter design adapted from \citep{wang_TES_2025} with permission from the authors. Right: simplified (not to scale) microcalorimeter model used in the simulation }
    \label{fig_TES}
\end{figure}

The microcalorimeter, which constitutes the detector array, was adapted from its original design (the left panel of Fig. \ref{fig_TES}; see also Fig. 1 in \cite{wang_TES_2025}). The microcalorimeter consists of an bi-layer absorber (Bi with the thickness of $5 ~\rm \mu m$ and Au with the thickness of $130 ~\rm n m$), a bi-layer TES (Mo and Cu) and supporting structures ($\rm Si_3N_4$ and Si). We simplified the architecture, but the essential geometry was preserved. The absorber stem and indentation were omitted, as their dimensions are much smaller than those of the absorbers. The Si substrate, whose cutaway view is trapezoidal in the original design, was approximated as a rectangular shape, since the difference between the lengths of the top and bottom surfaces is small compared to its height. The $\rm Si_3N_4$ substrate at the bottom of the microcalorimeter was also neglected due to its much smaller thickness relative to the two substrates above it. The simplified model is shown in the right panel of Fig. \ref{fig_TES}. These simplifications are expected to have a negligible impact on the simulated background because these structures are either very small in size or very thin. The probability of particle interactions with these components is minimal. The total effective area of the detector array is $1~\rm cm^2$. For components inside the detector assembly (DA) (e.g., the cold plate, Nb magnetic shield, Cryoperm shield) and the collimator filters, we implemented a high-fidelity geometrical description. For other components, minor structural details were omitted to reduce computational costs while retaining the bulk mass and material distribution.

The Bi and Au absorber layers on top of the microcalorimeter pixels were defined as the Sensitive Detectors in \textsc{Geant4} simulations. Energy depositions within these two volumes were recorded, and depositions falling within the DIXE energy band (0.1--10 keV) were treated as background events.

\subsection{Primary event Generation and normalization}\label{subsec: primary event generation}
Primary events were generated using the General Particle Source (GPS) in \textsc{Geant4}. In each simulation run, $N$ particles were emitted from a spherical surface of radius $R_{\rm ext}$, following a cosine-law angular distribution. The emission angle was restricted to $0 \leq \theta \leq \zeta$, forming a spherical envelope with radius $R_{\rm int} = R_{\rm ext} \sin \zeta$ that fully enclosed the mass model. This angular restriction suppresses particles propagating away from the payload, thereby improving computational efficiency.

The normalization from the simulated particle number $N$ to the physical background flux was performed following the method described in \cite{fioretti_SPIE_2012}. For a given particle population with an energy-integrated flux $\Phi$ (in units of $\rm counts ~cm^{-2} ~s^{-1} ~sr^{-1}$), if $C_i$ background counts are detected in energy bin $i$ with bin width $\Delta E_i$, and the detector has an effective area $A_{\rm det}$, the corresponding physical background flux $B_i$ (in $\rm counts~cm^{-2}~s^{-1}~keV^{-1}$) in energy bin $i$ is
\begin{equation}
    B_i = \frac{4\pi^2 C_i  R_{\rm int}^2 \Phi}{NA_{\rm det} \Delta E_i}
    \label{Bi}
\end{equation}

Because the Earth and its atmosphere partially block incident particles, some radiation components do not subtend a full $4\pi$ solid angle. We add a correction factor $f=\Omega_\mathrm{inc}/4\pi$ to Equation \eqref{Bi} to account for this effect \citep{galgoczi_configuration_2021}, yielding
\begin{equation}
    B_i^\prime = f B_i = \frac{\Omega_\mathrm{inc} \pi C_i  R_{\rm int}^2 \Phi}{N A_{\rm det} \Delta E_i}
    \label{Bi_prime}
\end{equation}
A detailed derivation of Equations \eqref{Bi} and \eqref{Bi_prime} is provided in Appendix \ref{appendix: event normalization}.

\subsection{Physics list} \label{subsec: physcis list}
\textsc{Geant4} is capable of simulating a wide range of physical processes governing particle-matter interactions. While some processes are essential for this work (e.g., ionization), others are irrelevant to the present application (e.g., quark-level interactions). The selection and configuration of the relevant processes are defined through the physics list.

In \cite{lotti_athena_2021}, the authors developed a custom electromagnetic physics list for simulating the particle background of the Athena\footnote{Athena \citep{nandra_athena_2013, barret_athena_2020} was a space mission proposed by the European Space Agency, focusing on the study of the hot and energetic universe. Later, this mission was re-desiged and became NewAthena \citep{cruise_newathena_2025}} X-IFU, which is also a TES-based microcalorimeter instrument. Since \textsc{Geant4} version 4.11.2, an example project derived from that work, named `xray\_TESdetector’\footnote{\url{https://geant4.kek.jp/lxr/source/examples/advanced/xray\_TESdetector/}}, has been included in the \textsc{Geant4} official example library. This example implements an electromagnetic physics list called `Space Physics List (SPL)', an updated version of the one used in \cite{lotti_athena_2021}, together with the hadronic physics list QBBC\footnote{\url{https://geant4-userdoc.web.cern.ch/UsersGuides/PhysicsListGuide/html/reference\_PL/QBBC.html}} \citep{ivantchenko_QBBC_2012}. Although originally optimized for Athena, SPL contains high-precision electromagnetic models suitable for more general space applications. The electromagnetic processes controlled by SPL include the photoelectric effect, Compton and Rayleigh scattering, bremsstrahlung, multiple scattering, ionization, pair production, and annihilation, with their energy range extended down to 100 eV.

Both (New)Athena X-IFU and DIXE employ TES-based microcalorimeters. Although the radiation environment at the Lagrange L1 point (the planned location for (New)Athena) differs from that in LEOs, the relevant physical processes for simulation are largely the same. We therefore adopted SPL as the electromagnetic physics list and QBBC as the hadronic physics list for our simulations. Particle transportation, decay, and radioactive decay processes are were also included.

Following \cite{lotti_athena_2021}, we divided the solids in the mass model into different regions according to their distance from the detector array and assigned different range cuts for different regions:
\begin{enumerate}
    \item The detector array and all solids directly visible from it (e.g., the cold plate, Nb magnetic shield, and the 4 K filter in the collimator) are assigned to the `inner region' with a range cut of $1 ~\rm \mu m$. For Bi, this range cut corresponds to an energy production cut of $1.11 ~\rm keV$ for photons and $5.30 ~\rm keV$ for electrons. For Au, this range cut corresponds to a production cut of $1.72 ~\rm keV$ for photons and $14.26 ~\rm keV$ for electrons. For Nb, this range cut corresponds to a production cut of $0.1 ~\rm keV$ for photons and $5.69 ~\rm keV$ for electrons. For Cu, this range cut corresponds to a production cut of $0.1 ~\rm keV$ for photons and $6.80 ~\rm keV$ for electrons.
    \item The remaining solids within the DA and the other filters in the collimator are assigned to the `intermediate region' with a range cut of $0.7 ~\rm mm$.
    \item All the other solids, including the supporting frames and the cooling system, are assigned to the `outer region' with a range cut of $1 ~\rm mm$, the default value in \textsc{Geant4}. For Al, this range cut corresponds to an energy production cut of $6.92 ~\rm keV$ for photons and $597.46 ~\rm keV$ for electrons.
\end{enumerate}

In Geant4, the range cut defines a range threshold: only induced particles capable of traveling farther than this distance are generated. For a given material, the range is internally converted into an equivalent energy threshold, i.e., the production cut. Assigning different range cuts to different regions allows for an effective balance between simulation accuracy and computational efficiency.

%%%%%%%%%%%%%%%%%%%%%%%%%%%%%%%%%%%%%%%%%%%%%%%%%%%%%%%%%%%%%%
\section{Results and discussion}\label{sec: results and discussion}

\subsection{NXB spectrum}\label{subsec: spectrum of NXB}

We first simulated the background at a geomagnetic latitude of $\theta_\mathrm{M}=0\degr$, an orbital altitude of $h=400 ~\rm km$, and during the solar minimum ($\phi=0.55 ~\rm GV$). This configuration represents a typical operating condition for DIXE, as most observations will be carried out at low geomagnetic latitudes. It also provided a conservative estimate since space radiation fluxes are higher during solar minimum, leading to an increased background level. The altitude of the CSS varies by only a few tens of kilometers, which is insufficient to significantly affect the particle flux or the resulting background. We therefore fix the orbital altitude at $h=400~\rm km$ throughout the following discussion unless stated otherwise.

\begin{figure}[t!]
    \centering
    \includegraphics[width=\textwidth]{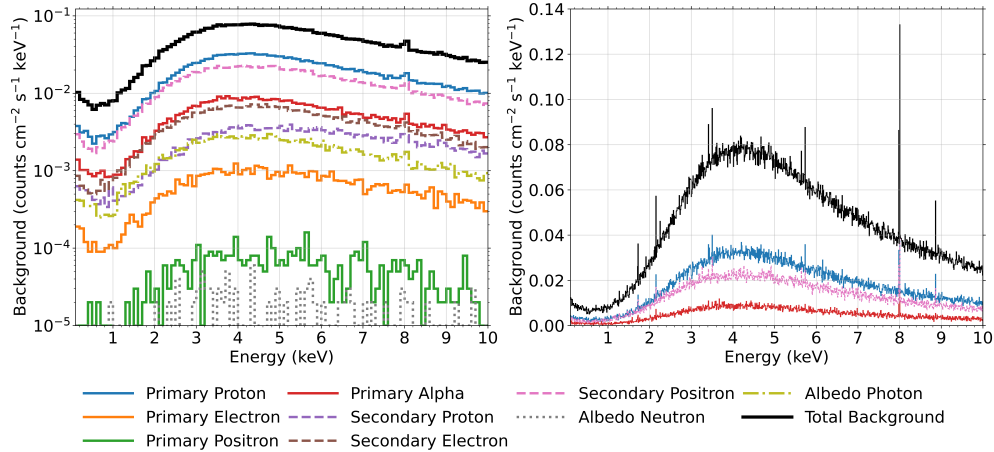}
    \caption{Spectrum of the NXB in the 0.1--10 keV band from different space radiation components at $\theta_\mathrm{M}=0\degr$ and $\phi=0.55 ~\rm GV$. Left: NXB spectrum from all space radiation particles with a  bin width of 0.1 keV.  The thick black line represents the total background obtained by summing contributions from all particle types. Right: high-resolution NXB spectrum (bin width of 10 eV) showing the three dominant components and the total background. Contributions from other components are not shown for clarity}
    \label{fig_NXB_spectrum}
\end{figure}

The average NXB within DIXE's energy band (0.1--10 keV), summed over all space radiation components, is $4.46 \times 10^{-2} ~\mathrm{counts~s^{-1}~cm^{-2}~keV^{-1}}$. The individual contributions from each component are listed in the second column of Table \ref{tab_bkg_latitude_1}, and the corresponding spectrum is shown in Fig. \ref{fig_NXB_spectrum}. Under these nominal conditions, the background is dominated by the background from primary protons, primary alpha particles, and secondary positrons. As shown in the left panel of Fig. \ref{fig_NXB_spectrum}, the combined continuum from all components decreases from $0.1$ to $ 1 ~\rm keV$, rises to a peak near 4 keV, and then subsequently declines toward 10 keV.

The mean energy loss due to ionization for charged particles traversing matter is described by the Bethe--Bloch equation:
\begin{equation}
    \left\langle-\frac{1}{\rho}\frac{d E}{d x}\right\rangle=K z^2 \frac{Z}{A} \frac{1}{\beta^2}\left[\frac{1}{2} \ln \frac{2 m_e c^2 \beta^2 \gamma^2 W_{\max }}{I^2}-\beta^2-\frac{\delta(\beta \gamma)}{2}\right]
    \label{bethe_bloch}
\end{equation}
Here $\frac{d E}{d x}$ is the energy loss per unit length, in units of $\rm MeV ~cm^{-1}$. $\rho$ is the density of the matter material, in units of $\rm g ~cm^{-3}$. $K=4 \pi N_A r_e^2 m_e c^2\approx 0.307 \rm ~MeV ~g^{-1} ~cm^2$ is a constant coefficient. $z$ is the charge number of the incident particle. $Z$ and $A$ are the atomic number and atomic mass of the matter material. The parameters $\beta=\frac{v}{c}$ and $\gamma$ denote the relativistic velocity and Lorentz factor of the incident particle. $W_{\max}$ is the maximum energy transfer in a single collision. $I$ is the mean excitation energy of the matter. $\delta$ is the density effect correction, which depends on $\beta \gamma.$ For more details of Equation \eqref{bethe_bloch}, we suggest that readers refer to \cite{navas_particle_physics_2024}. 

The cosmic particles typically behave as MIPs, i.e., the ionization per unit length along their path is comparable to the minimum value of Equation \eqref{bethe_bloch}. For Bi, $\left\langle-\frac{1}{\rho}\frac{d E}{d x}\right\rangle_{\min} \sim 1.1 \rm ~MeV ~g^{-1} ~cm^2$. Given the density $\rho \approx 9.8 ~\rm g ~cm^{-3}$ and the thickness $x = 5 ~\rm \mu m$ of the Bi absorber, the corresponding mean energy deposition is approximately $5.4 \rm ~keV$. 

The energy-loss distribution of MIPs in thin absorbers follows a Landau distribution, characterized by a peak and a long high-energy tail arising from rare large energy transfers. This behavior is consistent with the continuum spectrum shape in Fig.~\ref{fig_NXB_spectrum}. The most probable value of the Landau distribution is about $70\%$ of the mean energy loss, corresponding in this case to a peak of the spectrum at $\sim 4 \rm ~keV$, consistent with the peak seen in Fig.~\ref{fig_NXB_spectrum}.

The right panel of Fig. \ref{fig_NXB_spectrum} presents the NXB spectrum at a higher energy resolution. The number of simulation events used to produce this spectrum corresponds to an effective exposure time of $1 \rm ~Ms$ for DIXE. In addition to the continuum, the spectrum exhibits several discrete features. These emission lines are `instrument lines' and are produced by interactions between energetic particles and the instrument materials.

Here we note that the accuracy of \textsc{Geant4} electromagnetic models degrades progressively below $\sim 1 ~\rm keV$. Although we adopted a physics list specifically optimized for low-energy electromagnetic processes (see Section \ref{subsec: physcis list}), the reliability of models still becomes increasingly uncertain below 0.2 keV due to limitations in the underlying theoretical approximations, cross-section data, and the absence of detailed atomic and condensed-matter effects. Therefore, the spectrum below 0.2 keV should be treated with caution. However, in our simulation, the background events in 0.1--0.2 keV account for less than 1\% of all the background events in the 0.1--10 keV band. As a result, the limited accuracy of the physical models at very low energies does not have a significant impact on our overall conclusions.

\subsection{Characteristics of particles generating background} \label{Characteristics of particles}

\begin{figure}[t!]
    \centering
    \includegraphics[width=\textwidth]{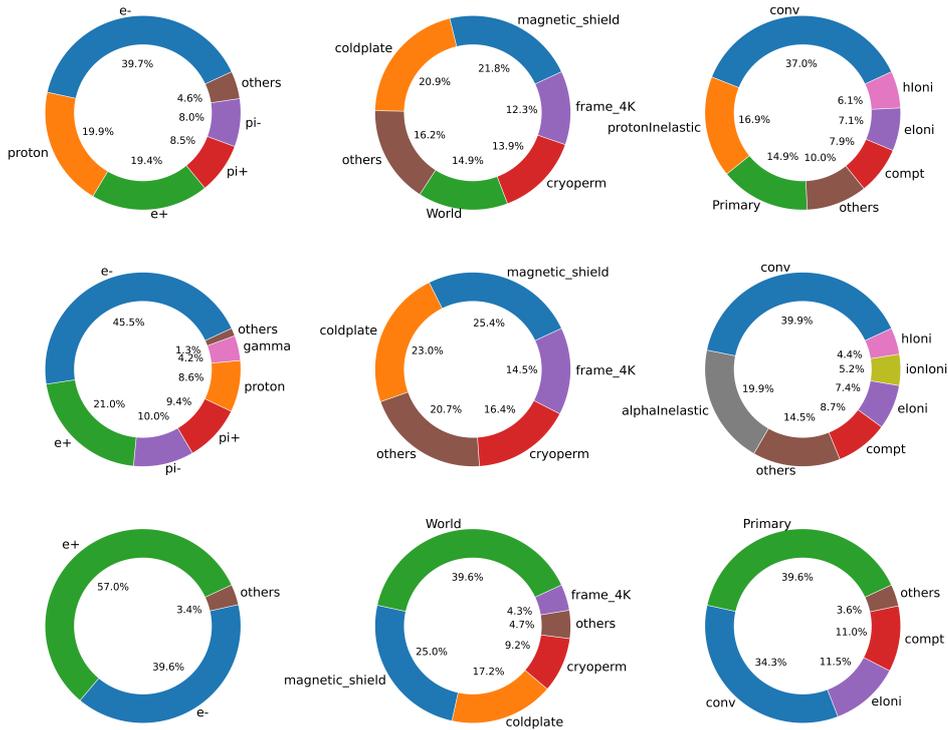}
    \caption{Relative contributions of particle types (left), creation volumes  (middle) and creation processes (right) of particles generating background events in the detectors. From top to bottom, the rows correspond to simulations with primary protons, primary alpha particles, and secondary positrons as the incident particles. Only contributions exceeding 4\% are shown explicitly; the others are grouped into `others'. The `coldplate', `magnetic\_shield' and `cryoperm' volume refers to the cold plate, Nb magnetic shield and Cryoperm shield of the DA. The `frame\_4K' volume refers to the Al support structure surrounding the DA and the ADR. The `world' volume denotes the vacuum region, and particles originating from this volume correspond to incident particles. The physical processes are labeled using \textsc{Geant4} terminology: `conv' refers to photon conversion (i.e., pair production); `compt' refers to Compton scattering; `eIoni', `hIoni' and `ionIoni' refers to electron, hadron and ion ionization; `protonInelastic', `neutronInelastic' and `alphaInelastic' refers to proton, neutron and alpha particle inelastic scattering; `Primary' means the particle is the incident cosmic particle}
    \label{fig_characteristics_three}
\end{figure}

To investigate the origin of the NXB, we tracked all particles depositing energy in the detector absorbers and recorded their particle types, creation locations, and production processes. This information allows us to identify which particles generate the background, where they are produced, and through what physical mechanisms they are produced. As an illustrative example, Fig.\ref{fig_characteristics_three} summarizes these properties for the background induced by primary protons, primary alpha particles, and secondary positrons under the nominal conditions of $\theta_\mathrm{M}=0^{\circ}$ and $\phi=0.55~\rm GV$.

For simulations with primary protons and primary alpha particles as incident radiation, the majority of the background originates from induced particles, with electrons providing the dominant contribution, as seen in the upper and middle panels of Fig. \ref{fig_characteristics_three}. In addition to the ionization process discussed in Section \ref{subsec: spectrum of NXB}, the GeV primary cosmic protons and cosmic alpha particles can undergo hadronic interactions (e.g., proton inelastic scattering, alpha particle inelastic scattering) with matter, producing hadronic showers, especially pions. These pions may either deposit energy directly in the absorber (see the contribution from pi+ and pi- in the upper left chart of Fig. \ref{fig_characteristics_three}), or decay into high-energy photons, which subsequently initiate electromagnetic showers: Photons convert into electron-positron pairs via pair production (`conv' processes in the right panel of Fig. \ref{fig_characteristics_three}), while electrons generate photons through Bremsstrahlung. The electromagnetic showers can also be triggered by GeV primary cosmic electrons and cosmic positrons. The induced particles produced through the processes mentioned above contribute significantly to the background. Given the short lifetime of pions, we examined the proper time (i.e., time considering the relativity effect) of pions directly depositing energy in the detectors and found it to be typically on the order of a few nanoseconds, well below their decay times. The induced particles responsible for the background are produced predominantly in structures close to the detector array, such as the magnetic shield, cold plate, and Cryoperm shield. Particles generated farther from the detectors are less likely to contribute, as they are either absorbed by intervening material or propagate away from the detector volume.

In contrast, for secondary positrons, nearly 40\% of the background is produced directly by the incident particles themselves, as indicated by the `world' volume and the `Primary' creation process in the bottom panels in Fig. \ref{fig_characteristics_three}. A similar trend is also observed for secondary protons and electrons. The energies of secondary cosmic particles are typically at the MeV scale, which is insufficient to efficiently trigger hadronic or electromagnetic showers, unlike primary cosmic particles. Consequently, they can only directly deposit energy in the absorbers or produce induced particles through ionization. Because these secondary particles are less effective at generating additional induced
particles, their background contribution is more directly associated with the incident particles themselves, and the relative importance of induced particles is lower compared to the case of primary cosmic rays.

These results indicate that DIXE's design provides effective shielding against high-energy primary cosmic rays while simultaneously producing a substantial population of induced particles in the surrounding structures. This trade-off between shielding efficiency and induced background is a key factor in determining the overall NXB level for TES-based instruments in LEOs.

\subsection{Background at different geomagnetic latitudes} \label{subsec: bkg at different latitudes}
As discussed in Section \ref{sec: space radiation environment}, the space radiation environment in LEOs varies strongly with geomagnetic latitude. To quantify its impact on the instrumental background, we simulated the NXB produced by each space radiation component at different geomagnetic latitudes $\theta_\mathrm{M}$, sampled in steps of $10\degr$ up to $50\degr$, with an additional simulation at $\theta_\mathrm{M}=53\degr$, corresponding to the maximum latitude reached by the CSS orbit. The number of simulation events for each $\theta_\mathrm{M}$ and radiation component corresponds to an effective exposure time of $100 \rm ~ks$, except for $\theta_\mathrm{M}=0\degr$ where $1 \rm ~Ms$ applies. 

\begin{figure}[t!]
    \centering
    \includegraphics[width=\textwidth]{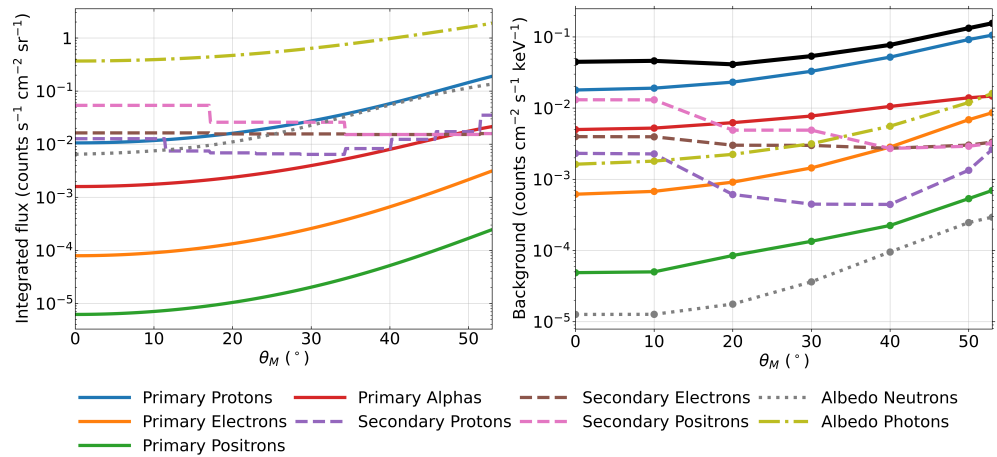}
    \caption{Integrated flux of different space radiation components (left) and  background from these components at different geomagnetic latitudes $\theta_\mathrm{M}$ (right)}
    \label{fig_bkg_latitude}
\end{figure}

The resulting background levels are listed in Table \ref{tab_bkg_latitude_1} and Table \ref{tab_bkg_latitude_2}, and are shown in the right panel of Fig. \ref{fig_bkg_latitude}. For each radiation component, the variation of the background with geomagnetic latitude closely follows the change in the corresponding integrated cosmic particle flux (left panel of Fig. \ref{fig_bkg_latitude}). For the total background, it remains nearly constant for $\theta_\mathrm{M}\leq20\degr$, but increases rapidly at higher geomagnetic latitudes.

\begin{table}[t!]
\caption {Background at $\theta_\mathrm{M} = 0\degr, 10\degr, 20\degr$ for $\phi=0.55 ~\rm GV$ and $h=400 ~\rm km$. The background is in units of $\mathrm{counts ~s^{-1} ~cm^{-2} ~keV^{-1}}$}
\label{tab_bkg_latitude_1}
\centering
\begin{tabular}{lccc}
    \toprule
    $\theta_\mathrm{M}$ & $0\degr$ & $10\degr$ & $20\degr$ \\
    \midrule
    Primary proton & $(1.79\pm 0.01)\times 10^{-2}$ & $(1.90\pm 0.01)\times 10^{-2}$ & $(2.31 \pm 0.02) \times 10^{-2}$ \\
    Primary electron & $(6.16\pm 0.08)\times 10^{-4}$ & $(6.76\pm 0.26)\times 10^{-4}$ & $(9.10 \pm 0.30) \times 10^{-4}$ \\
    Primary positron & $(4.90\pm 0.22)\times 10^{-5}$ & $(5.00\pm 0.71)\times 10^{-5}$ & $(8.48 \pm 0.93) \times 10^{-5}$ \\
    Primary alpha & $(4.99\pm 0.02)\times 10^{-3}$ & $(5.23\pm 0.07)\times 10^{-3}$ & $(6.23\pm 0.08) \times 10^{-3}$\\
    Secondary proton & $(2.31\pm 0.02)\times 10^{-3}$ & $(2.27\pm0.05)\times 10^{-3}$ & $(6.11\pm 0.25) \times  10^{-4}$\\
    Secondary electron & $(3.97\pm 0.02)\times 10^{-3}$ & $(3.95\pm0.06)\times 10^{-3}$ & $(3.01 \pm 0.05)\times 10^{-3}$ \\
    Secondary positron & $(1.31\pm 0.01)\times 10^{-2}$ & $(1.30\pm0.01)\times 10^{-2}$ & $(4.89 \pm 0.07) \times 10^{-3}$\\
    Albedo neutron & $(1.26\pm 0.11)\times 10^{-5}$ & $(1.27\pm0.34)\times 10^{-5}$ & $(1.76 \pm 0.42) \times 10^{-5}$\\
    Albedo photon & $(1.63\pm 0.01)\times 10^{-3}$ & $(1.80\pm0.04)\times 10^{-3}$ & $(2.24 \pm 0.05) \times 10^{-3}$\\
    \midrule
    \textbf{Total} & $(4.46\pm0.01) \times 10^{-2}$ & $(4.60\pm0.02) \times 10^{-2}$ & $(4.11\pm0.02) \times 10^{-2}$ \\
    \botrule
\end{tabular}
\end{table}

\begin{table}[t!]
\caption {Background at $\theta_\mathrm{M} = 30\degr, 40\degr, 50\degr, 53\degr$ for $\phi=0.55 ~\rm GV$ and $h=400 ~\rm km$. The background is in units of $\mathrm{counts ~s^{-1} ~cm^{-2} ~keV^{-1}}$}
\label{tab_bkg_latitude_2}
\centering
\begin{tabular}{lcccc}
    \toprule
    $\theta_\mathrm{M}$ & $30\degr$ & $40\degr$ & $50\degr$ & $53\degr$\\
    \midrule
    Primary proton & $(3.28 \pm0.02) \times 10^{-2}$ & $(5.19 \pm 0.02) \times 10^{-2}$ & $(9.17\pm 0.03)\times10^{-2}$ & $(1.06\pm 0.01)\times10^{-1}$\\
    Primary electron & $(1.44 \pm 0.04) \times 10^{-3}$ & $(2.83 \pm 0.05) \times 10^{-3}$ & $(6.85\pm 0.07)\times10^{-3}$ & $(8.60\pm 0.09)\times10^{-3}$\\
    Primary positron & $(1.34 \pm 0.12) \times 10^{-4}$ & $(2.24\pm 0.15) \times 10^{-4}$ & $(5.34\pm 0.23)\times10^{-4}$ & $(6.96\pm 0.26)\times10^{-4}$\\
    Primary alpha & $(7.73 \pm 0.08) \times 10^{-3}$ & $(1.05 \pm 0.01) \times 10^{-2}$ & $(1.39\pm 0.01)\times10^{-2}$ & $(1.47\pm 0.01)\times10^{-2}$\\
    Secondary proton & $(4.47 \pm 0.21) \times 10^{-4}$ & $(4.43 \pm 0.21) \times 10^{-4}$ & $(1.34\pm 0.04)\times10^{-3}$ & $(2.63\pm 0.05)\times10^{-3}$\\
    Secondary electron & $(2.99 \pm 0.05) \times 10^{-3}$ & $(2.72 \pm 0.05) \times 10^{-3}$ & $(3.02\pm 0.05)\times10^{-3}$ & $(3.32\pm 0.06)\times10^{-3}$\\
    Secondary positron & $(4.89 \pm 0.07) \times 10^{-3}$ & $(2.71 \pm 0.05) \times 10^{-3}$ & $(2.91\pm 0.05)\times10^{-3}$ & $(3.20\pm 0.06)\times10^{-3}$\\
    Albedo neutron & $(3.62 \pm 0.60) \times 10^{-5}$ & $(9.52 \pm 0.98) \times 10^{-5}$ & $(2.46\pm 0.16)\times10^{-4}$ & $(2.94\pm 0.17)\times10^{-4}$ \\
    Albedo photon & $(3.15 \pm 0.06) \times 10^{-3}$ & $(5.56 \pm 0.07) \times 10^{-3}$ & $(1.20\pm 0.01)\times10^{-2}$ & $(1.61\pm 0.01)\times10^{-2}$\\
    \midrule
    \textbf{Total} & $(5.37\pm0.02) \times 10^{-2}$ & $(7.70\pm0.03) \times 10^{-2}$ & $(1.33\pm0.04) \times 10^{-1}$ & $(1.55\pm0.04) \times 10^{-1}$ \\
    \botrule
\end{tabular}
\end{table}

At low geomagnetic latitudes, the total background is dominated by contributions from primary protons, primary alpha particles, and secondary positrons. With increasing $\theta_\mathrm{M}$, the background induced by primary protons and alpha particles rises steadily as the geomagnetic cut-off rigidity decreases, while the contribution from secondary positrons decreases. Up to $\theta_\mathrm{M}\sim20\degr$, these opposite trends largely compensate for each other, resulting in a relatively stable total background. At higher geomagnetic latitudes, however, the decrease in the secondary positron component is no longer sufficient to offset the rapidly increasing background from primary protons. As a result, the total background becomes dominated by primary cosmic protons and increases sharply with $\theta_\mathrm{M}$, reaching $1.55 \times 10^{-1} ~\mathrm{counts~s^{-1}~cm^{-2}~keV^{-1}}$ at $\theta_\mathrm{M}=53\degr$, approximately four times the value at $\theta_\mathrm{M}=0\degr$.

These results clearly demonstrate that the geomagnetic latitude is a key factor governing the NXB level in LEOs. From the perspective of background minimization, a low-inclination orbit ($\theta_\mathrm{M}\lesssim30\degr$) is therefore strongly preferred for soft X-ray observation missions.

\subsection{Background for different solar activities} \label{bkg at different solar activities}
To assess the impact of solar activity on the NXB, we performed simulations for solar minimum ($\phi=0.55~\rm GV$) and solar maximum ($\phi=1.1~\rm GV$) conditions at two representative geomagnetic latitudes, $\theta_\mathrm{M}=0\degr$ and $\theta_\mathrm{M}=53\degr$. The results are summarized in Table \ref{tab_bkg_solar}. As discussed in Section \ref{subsubsec: secondary cosmic particles}, the flux of secondary cosmic particles does not change with solar activity, so the background from secondary cosmic particles remains unchanged between solar minimum and solar maximum. In contrast, the fluxes of primary cosmic rays are modulated by solar activity, leading to a reduction in their corresponding background contributions during solar maximum.

\begin{table}[t!]
    \centering
    \caption{Background for different solar activities and geomagnetic latitudes. The background is in units of $\mathrm{counts ~s^{-1} ~cm^{-2} ~keV^{-1}}$}
    \label{tab_bkg_solar}%
    \begin{tabular}{lcccc}
    \toprule
    $\phi$ & 0.55 GV & 1.1 GV & 0.55 GV & 1.1 GV \\
    $\theta_\mathrm{M}$ & $0\degr$ & $0\degr$ & $53\degr$ & $53\degr$ \\
    \midrule
    Primary proton & $(1.79\pm 0.01)\times 10^{-2}$ & $(1.62\pm 0.01)\times 10^{-2}$ & $(1.06\pm 0.01)\times10^{-1}$ & $(6.54\pm 0.03)\times 10^{-2}$\\
    Primary electron & $(6.16\pm 0.08)\times 10^{-4}$ & $(5.32\pm 0.23)\times 10^{-4}$ & $(8.60\pm 0.09)\times10^{-3}$ & $(4.52\pm 0.06)\times 10^{-4}$\\
    Primary positron & $(4.90\pm 0.22)\times 10^{-5}$ & $(4.42\pm 0.68)\times 10^{-5}$ & $(6.96\pm 0.27)\times10^{-4}$ & $(3.51\pm 0.19)\times10^{-4}$\\
    Primary alpha & $(4.99\pm 0.02)\times 10^{-3}$ & $(4.45\pm 0.06)\times 10^{-3}$ & $(1.47\pm 0.01)\times10^{-2}$ & $(1.04\pm 0.01)\times10^{-2}$\\
    Secondary proton & $(2.31\pm 0.02)\times 10^{-3}$ & $(2.26\pm 0.05)\times 10^{-3}$ & $(2.63\pm 0.05)\times10^{-3}$ & $(2.63\pm 0.05)\times10^{-3}$\\
    Secondary electron & $(3.97\pm 0.02)\times 10^{-3}$ & $(3.95\pm 0.06)\times 10^{-3}$ & $(3.32\pm 0.06)\times10^{-3}$ & $(3.29\pm 0.06)\times10^{-2}$\\
    Secondary positron & $(1.31\pm 0.01)\times 10^{-2}$ & $(1.31\pm 0.01)\times 10^{-2}$ & $(3.20\pm 0.06)\times10^{-3}$ & $(3.32\pm 0.06)\times10^{-3}$\\
    Albedo neutron & $(1.26\pm 0.11)\times 10^{-5}$ & $(1.05\pm 0.30)\times 10^{-5}$ & $(2.94\pm 0.17)\times10^{-4}$ & $(1.44\pm 0.12)\times10^{-4}$\\
    Albedo photon & $(1.63\pm 0.01)\times 10^{-3}$ & $(1.73\pm 0.04)\times 10^{-3}$ & $(1.61\pm 0.01)\times10^{-2}$ & $(1.60\pm 0.01)\times10^{-2}$\\
    \midrule
    \textbf{Total} & $(4.46 \pm 0.01) \times 10^{-2}$ & $(4.23 \pm 0.02) \times 10^{-2}$ & $(1.55 \pm 0.02) \times 10^{-1}$ & $(1.06 \pm 0.02) \times 10^{-1}$\\
    \botrule
    \end{tabular}%
\end{table}

At low geomagnetic latitudes, the transition from solar minimum to solar maximum results in only a modest decrease in the total background, by approximately 5\%. At high geomagnetic latitudes ($\theta_\mathrm{M}=53\degr$), the same change in solar activity leads to a substantially larger reduction in the total background, of about 30\%.  As discussed in Section \ref{subsubsec: primary cosmic particles} and shown in Fig. \ref{fig_p_p_spectrum}, variations in solar activity lead to only modest changes in the flux of space radiation, especially at low geomagnetic latitudes. Therefore, the effect of solar activity on the background is small compared to that of geomagnetic latitude. 

Overall, the influence of solar activity on the NXB is secondary compared to that of geomagnetic latitude. Nevertheless, solar modulation introduces long-term variations in the background level that cannot be neglected. Since DIXE will primarily operate in a sky-survey mode, such temporal variations may result in differences in data quality for the same sky region observed at different epochs.

\subsection{Delayed background from the SAA}

\begin{figure}[t!]
    \centering
    \includegraphics[width=0.7\textwidth]{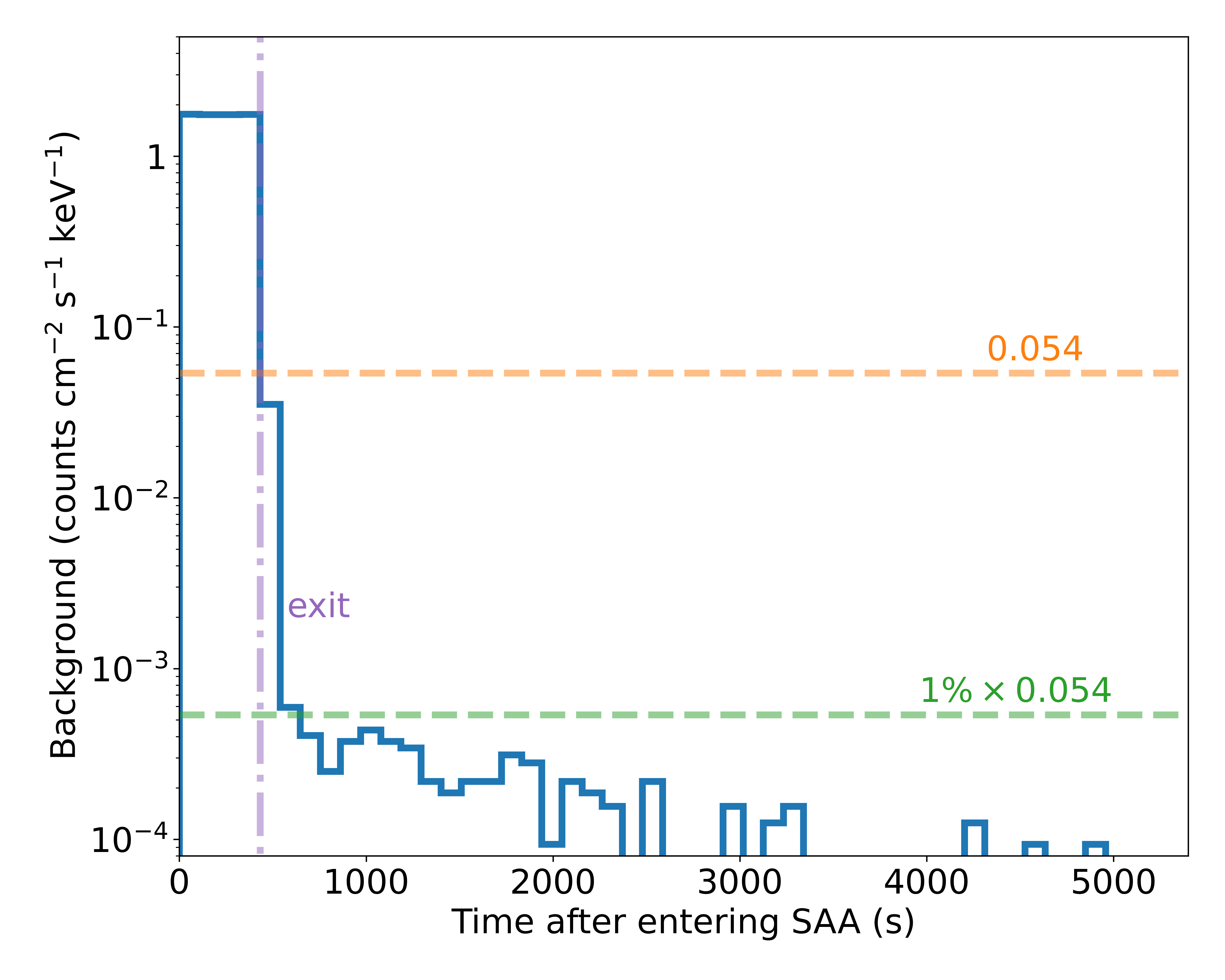}
    \caption{Background variation as a function of time after entering the SAA. The blue solid histogram shows the background level over time. The orange and green dashed lines correspond to 100\% and 1\% of the prompt background at $\theta_\mathrm{M} = 30\degr$, respectively. The purple vertical dot-dashed line marks the time of exiting the SAA}
    \label{fig_delayed_bkg}
\end{figure}

After passing through the SAA, activated materials in the payload continue to emit particles via radioactive decay, producing a delayed background component. We simulated the delayed NXB induced by trapped protons during SAA passages. Fig.~\ref{fig_delayed_bkg} illustrates the temporal evolution of this delayed background following entry into the SAA.

While the spacecraft is inside the SAA, the background from trapped protons is extremely high, reaching $1.76 \rm ~counts ~s^{-1} ~cm^{-2} ~keV^{-1}$. Immediately after leaving the SAA, the delayed background drops rapidly to $4\times10^{-2} \rm ~counts ~s^{-1} ~cm^{-2} ~keV^{-1}$, close to the prompt background level at a geomagnetic latitude comparable to that of the SAA ($\theta_\mathrm{M}\approx30\degr$). After approximately 5 minutes, the delayed background decreases to only about 1\% of the prompt background.

In summary, the delayed background induced by trapped protons in the SAA has a noticeable impact only during the first few minutes after exiting the anomaly. At later times, its contribution becomes negligible compared to the prompt background. Consequently, the influence of SAA passages on normal scientific observations is limited, provided that data acquired shortly after SAA exit are appropriately filtered. 

\subsection{Comparison with other soft X-ray telescopes}

In Table \ref{tab_comparison_telescope}, we compare the simulated NXB of DIXE with those of other launched or proposed soft X-ray microcalorimeter-based telescopes, including Suzaku/XRS \citep{mitsuda_suzaku_2007, kelley_suzaku_2007}, Hitomi/SXS \citep{Kelley_hitomi_2016, kilbourne_hitomi_2018}, and Athena/X-IFU \citep{lotti_athena_2021}. The quantity labeled `NXB without ACD' refers to the unscreened background from cosmic particles prior to the application of an ACD. An exception is Hitomi/SXS, for which the quoted value corresponds to the raw in-orbit background, including frame events and other instrumental contributions. The `NXB with ACD' denotes the residual background after applying ACD rejection. 

The simulated NXB level of DIXE without ACD is very close to that of Suzaku/XRS and Hitomi/SXS, indicating that our simulation provides a realistic estimate of the particle-induced background. Although DIXE will operate in LEOs, whereas (New)Athena/X-IFU is to be located at the L1 point, we can use the DIXE background at $\theta_M=53\degr$ for a qualitative comparison. At high geomagnetic latitudes, the modulation effect of the Earth's geomagnetic field is weak, so the cosmic particle flux is similar to that at L1 point. The background of DIXE at $\theta_M=53\degr$ is $1.55 \times 10^{-1} ~\mathrm{counts~s^{-1}~cm^{-2}~keV^{-1}}$, very close to the corresponding background of Athena/X-IFU. This again confirms the reliability of our work. For Suzaku, Hitomi, and Athena, both simulations and in-orbit measurements demonstrate that the implementation of an ACD can significantly suppress the NXB. For DIXE, however, incorporating an ACD is not currently planned. Despite the relatively higher background, the large FOV, the sky-survey observing strategy, and the high energy resolution will enable DIXE to achieve its planned scientific objectives (Liu et al. in prep.). 

\begin{table}[t!]
    \centering
    \caption{Comparison of the NXB levels of DIXE with Suzaku, Hitomi and Athena. The background is in units of $\mathrm{counts ~s^{-1} ~cm^{-2} ~keV^{-1}}$. The NXB data of other telescopes are taken from \cite{kelley_suzaku_2007, Kelley_hitomi_2016, kilbourne_hitomi_2018, lotti_athena_2021}}
    \label{tab_comparison_telescope}
    \begin{tabular}{lcccc}
    \toprule
    Instrument & DIXE & Suzaku/XRS & Hitomi/SXS & Athena/X-IFU \\
    \midrule
    Orbit & LEO (400 km) & LEO (570 km) & LEO (575 km) & L1 \\
    Energy band & 0.1--10 keV & 0.1--12 keV & 0.3--12 keV & 2--10 keV \\
    Data type & simulation & observation & observation & simulation \\
    NXB without ACD & $4.5 \times 10^{-2}$ & $4.6 \times 10^{-2}$ & $4.7 \times 10^{-2}$ & $1.5 \times 10^{-1}$ \\
    NXB with ACD & / & $4.2 \times 10^{-3}$ & $3.4 \times 10^{-3}$ & $4.8 \times 10^{-3}$ \\
    \botrule
    \end{tabular}
\end{table}

We further compared the unscreened NXB spectrum of DIXE (Fig. \ref{fig_NXB_spectrum}) with those of Suzaku (Fig. 44 in \cite{kelley_suzaku_2007}), Hitomi (Fig. 9 in \cite{kilbourne_hitomi_2018}), and Athena (Fig. 3 in \cite{lotti_athena_2021}). The NXB spectra of Hitomi/SXS and Athena/X-IFU closely resemble that of DIXE, exhibiting a decline from low energies to $\sim 2 ~ \rm keV$, a rise toward a broad peak near $\sim 5 ~ \rm keV$, and a subsequent decrease toward higher energies. The spectral resolution of Suzaku/XRS is insufficient for a detailed shape comparison.

%%%%%%%%%%%%%%%%%%%%%%%%%%%%%%%%%%%%%%%%%%%%%%%%%%%%%%%%%%%%%%
\section{Conclusions}\label{sec: conclusion}

In this work, we performed a comprehensive simulation of the NXB for DIXE using the \textsc{Geant4} toolkit. A detailed mass model of the instrument payload was constructed, and all major space radiation components -- including primary and secondary cosmic rays, albedo neutrons, and albedo photons -- were modeled with realistic spatial and spectral distributions. The simulations incorporated appropriate event generation and normalization procedures, together with the Space Physics List (SPL), to reproduce the physical radiation environment in LEOs.

Our results show that the typical NXB count rate is $4.46 \times 10^{-2} ~\mathrm{counts~s^{-1}~cm^{-2}~keV^{-1}}$ at a geomagnetic latitude $\theta_\mathrm{M} = 0\degr$ during the solar minimum. The dominant contributors to the NXB are primary cosmic protons. Most background is generated by induced particles, particularly electrons, produced in materials close to the detector array, with the Nb magnetic shield playing a major role. The NXB level exhibits a strong dependence on geomagnetic latitude, increasing significantly toward higher latitudes as a result of the reduced geomagnetic COR, and reaching $1.55 \times 10^{-1} ~\mathrm{counts~s^{-1}~cm^{-2}~keV^{-1}}$ at $\theta_\mathrm{M}=53\degr$. Variations in the solar cycle were also investigated, showing that the background level is higher during solar minimum than during solar maximum, owing to weaker solar modulation of cosmic rays.

We also studied the delayed background induced by trapped protons in the SAA. Although the background level inside the SAA is extremely high, it decreases rapidly after exiting the anomaly. The delayed component becomes negligible within approximately 5 minutes, indicating that its impact on routine scientific observations is confined to a very short interval following SAA passages.

Finally, a comparison with other soft X-ray microcalorimeter missions demonstrates that the simulated NXB level of DIXE is consistent with expectations for instruments of similar design and orbital parameters. These results support the feasibility of the DIXE payload design. Our analyzes provide insights into the background of microcalorimeter-based X-ray telescopes. The analysis presented here provides insights into the particle-induced background of microcalorimeter-based X-ray instruments, and the simulation framework developed in this work can be readily extended to optimize instrument design, refine observational strategies, and support future end-to-end observation simulations.

\backmatter

\bmhead{Acknowledgments}
We sincerely thank Fei Xie and Zuke Feng for their generous support at the early stage of this project. Their assistance in resolving several \textsc{Geant4}-related questions was essential to the completion of this study. We also thank Jiahuan Zhu, Yuning Zhang, Chunyang Jiang and Naihui Chen for helpful discussions and valuable suggestions. We gratefully thank the anonymous referee for constructive feedback, which has greatly improved the quality and clarity of this article. This work was supported by the National Natural Science Foundation of China (Grant No. 12220101004, 11821303) and the Ministry of Science and Technology of China (Grant No. 2018YFA0404502). J.M. acknowledges support from the Tsinghua Dushi Program 53121200125. We further acknowledge the Tsinghua Astrophysics High-Performance-Computing (TAHPC) platform for providing computational and data storage resources.

\bmhead{Author Contributions}
R.T. carried out the \textsc{Geant4} simulations, performed data analysis, and drafted the manuscript. J.M. and W.C. provided scientific guidance throughout the study. J.L. and H.J. designed the instrument and provided the design as the mass model used in the simulations. W.C. supervised the project and revised the manuscript. All authors reviewed the manuscript.

\bmhead{Funding}
This work was supported by the National Natural Science Foundation of China (Grant No. 12220101004, 11821303) and the Ministry of Science and Technology of the People's Republic of China (Grant No. 2018YFA0404502). Junjie Mao acknowledges support from the Tsinghua Dushi Program 53121200125.

\bmhead{Data availability}
All data are available from the corresponding author upon reasonable request.

\section*{Declarations}

\bmhead{Competing interests}
The authors declare no competing interests.

\begin{appendices}
\section{Event normalization procedure}\label{appendix: event normalization}
For a cosmic particle population with the energy-integrated flux $\Phi$ (in units of $\rm counts ~cm^{-2} ~s^{-1} ~sr^{-1}$), if $N$ particles are emitted from the spherical surface of radius $R_\mathrm{ext}$ following a cosine-law angular distribution, the emission rate $P$ (in units of $\rm counts ~s^{-1}$) is given by:
\begin{equation}
    P = \Phi \times A_{\rm ext} \times \Omega_{\mathrm{em}}
\end{equation}
where $A_{\rm ext} = 4\pi R_{\rm ext}^2$ is the surface area of the source sphere, and $\Omega_{\rm em}$ is the solid angle of emission. For a cosine-law distribution with an angular restriction $0\leq\theta\leq\zeta$, the emission solid angle is 
\begin{align}
    \Omega_{\rm em} &= \int_\theta \int_\phi \cos \theta \sin \theta ~\mathrm{d} \theta ~\mathrm{d} \phi \nonumber \\
    &= \int_0^\zeta \cos \theta \sin \theta ~\mathrm{d} \theta \int_0^{2\pi} ~\mathrm{d} \phi~ = \pi \sin^2{\zeta}
\end{align}
Substituting $\Omega_{\rm em}$ into the emission rate $P$ yields
\begin{equation}
    P = \Phi \times 4\pi R_{\rm ext}^2 \times \pi \sin^2\zeta = 4\pi^2 R_{\rm int}^2 \Phi
\end{equation}

The real-world exposure time $\tau$ corresponding to this simulation run with $N$ emitted particles is therefore
\begin{equation}
    \tau = \frac{N}{P} = \frac{N}{4\pi^2 R_{\rm int}^2 \Phi}
\end{equation}

If $C_i$ background counts are detected in energy bin $i$ with width $\Delta E_i$, and the detector has an effective area $A_{\rm det}$, the physical background flux $B_i$ (Equation \eqref{Bi}, in units of $\rm counts~cm^{-2}~s^{-1}~keV^{-1}$) in energy bin $i$ is
\begin{equation}
    B_i = \frac{C_i}{\tau A_{\rm det} \Delta E_i} = \frac{4\pi^2 C_i  R_{\rm int}^2 \Phi}{NA_{\rm det} \Delta E_i}
\end{equation}

Because the orbital altitude of LEO is small compared to the Earth's radius,a substantial fraction of the sky is shadowed by the Earth and its atmosphere. The visible sky fraction can be described by the horizon angle $\theta_{\rm H}$, defined as the angle between the zenith and the top of the atmosphere \citep{cumani_background_2019}:
\begin{equation}
    \theta_{\rm H}=90^{\circ}+\arccos \frac{R_{\oplus}+H_{\rm A}}{R_{\oplus}+h}
\end{equation}
where $R_\oplus = 6371 \rm ~km$ is the Earth's radius, $H_{\rm A} = 40 \rm ~km$ is the atmospheric top altitude, and $h$ is the orbital altitude. For CSS, the orbital altitude is $h = 400 \rm ~km$, so the horizon angle is $\theta_{\rm H}=108^\circ$.

The correction factor $f=\Omega_\mathrm{inc}/4\pi$ is introduced in Equation \eqref{Bi_prime} to account for the Earth occultation effect  \citep{galgoczi_configuration_2021}:
\begin{equation}
    B_i^\prime = f B_i = \frac{\Omega_\mathrm{inc} \pi C_i  R_{\rm int}^2 \Phi}{N A_{\rm det} \Delta E_i}
\end{equation}

The primary cosmic particles originate only from the visible sky, so the incident solid angle is:
\begin{equation}
    \Omega_\mathrm{inc} = \int_\theta \int_\phi \sin \theta ~\mathrm{d} \theta ~\mathrm{d} \phi = \int_0^{\theta_\mathrm{H}} \sin \theta ~\mathrm{d} \theta \int_0^{2\pi} ~\mathrm{d} \phi = 2\pi (1-\cos \theta_\mathrm{H})
\end{equation}

The secondary cosmic particles and trapped particles inside the SAA are assumed to be isotropic, with $\Omega_\mathrm{inc} = 4\pi$ .

Albedo neutrons and photons originate from the atmosphere below the orbit, giving
\begin{equation}
    \Omega_\mathrm{inc} = \int_\theta \int_\phi \sin \theta ~\mathrm{d} \theta ~\mathrm{d} \phi = \int_{\theta_\mathrm{H}}^{2\pi} \sin \theta ~\mathrm{d} \theta \int_0^{2\pi} ~\mathrm{d} \phi = 2\pi (1+\cos \theta_\mathrm{H})
\end{equation}

\end{appendices}

\bibliography{sn-bibliography}% common bib file

\end{document}